\documentclass[aps,prd,superscriptaddress,floatfix,nofootinbib,notitlepage,12pt]{revtex4-1}

\usepackage{aas_macros}
\usepackage{graphicx}
\usepackage{multirow}
\usepackage{color}
\usepackage{slashed}
\usepackage{comment}
\usepackage{ulem}
\usepackage{soul}
\setstcolor{red}
\usepackage{amsmath,amssymb,amsfonts}
\usepackage{verbatim}
\usepackage{bm}
\usepackage{color}
\usepackage{ulem}
\usepackage{mathtools}
\usepackage{colortbl}
\usepackage[dvipsnames]{xcolor}
\usepackage{slashed}
\usepackage{braket}
\usepackage{adjustbox}
\usepackage{cancel}

\usepackage{ulem}
\usepackage{soul}
\setstcolor{red}

\usepackage[colorlinks]{hyperref}
\hypersetup{
	colorlinks=true,
	linkcolor=blue,
	filecolor=blue,      
	urlcolor=blue,
	citecolor=magenta,
	pdfpagemode=FullScreen
} 

\newcommand{\orcid}[1]{\hspace{1mm}\href{https://orcid.org/#1}{\includegraphics[height=0.3cm,keepaspectratio]{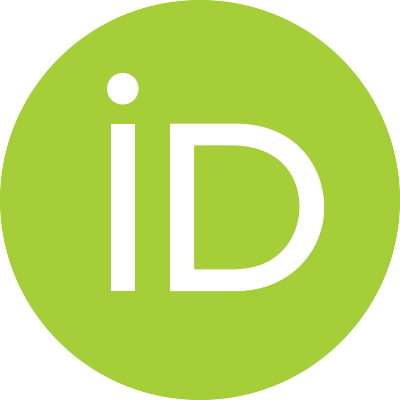}}}

\newcommand{\fb}{{\it fb}\,}
\newcommand{\br}{\text{Br}}

\newcommand{\hc}{\text{h.c}}

\begin{document}
	
\title{Associated charged Higgs production within the 2HDM:\\
$e^-e^+$ versus $\mu^-\mu^+$ colliders}
\author{Brahim Ait Ouazghour\orcid{0009-0006-1419-969X}}
\email{b.ouazghour@gmail.com}
\affiliation{LPHEA, FSSM, Cadi Ayyad University, P.O.B. 2390 Marrakech, Morocco}
\author{Abdesslam Arhrib\orcid{0000-0001-5619-7189}}
\email{aarhrib@gmail.com}
\affiliation{Abdelmalek Essaadi University, FST Tanger B.P. 416, Morocco}
\affiliation{Department of Physics and CTC, National Tsing Hua University, Hsinchu, Taiwan 300}
\author{Kingman Cheung\orcid{0000-0003-2176-4053}}
\email{cheung@phys.nthu.edu.tw}
\affiliation{Department of Physics and CTC, National Tsing Hua University, Hsinchu, Taiwan 300}
\affiliation{Division of Quantum Phases and Devices, School oproduction f Physics,
Konkuk University, Seoul 143-701, Republic of Korea}
\author{Es-said Ghourmin\orcid{0009-0007-1597-8537}}
\email{s.ghourmin123@gmail.com}
\affiliation{Laboratory of Theoretical and High Energy Physics (LPTHE), Faculty of Science, Ibnou Zohr University, B.P 8106, Agadir, Morocco}
\author{Larbi Rahili\orcid{0000-0002-1164-1095}}
\email{rahililarbi@gmail.com}
\affiliation{Laboratory of Theoretical and High Energy Physics (LPTHE), Faculty of Science, Ibnou Zohr University, B.P 8106, Agadir, Morocco}	
\date{\today}
\begin{abstract}
Our goal is to investigate the charged Higgs phenomenology in the framework of 2HDM at the upcoming $e^+e^-$ and muon colliders. We are primarily concerned with the associated production processes with a fermion pair: $\ell^+ \ell^- \to \tau^+ \nu_{\tau }H^-$ and $\ell^+ \ell^- \to t \bar{b} H^-$, as well as with the W boson and a neutral Higgs boson: $\ell^+ \ell^- \to W^\pm H^\mp S$ ($S=h,\,H,\, A$) and $\ell^+ \ell^- \to W^\pm H^\mp Z$. We first update the results for $e^+e^- \to \{ \tau^+ \nu_{\tau }H^-\ , \  t \bar{b} H^-\}$  and then discuss our findings for $e^+e^- \to \{ W^\pm H^\mp S\}$  for various center of mass energies 500 GeV, 1 TeV, 1.5 TeV and 3 TeV. In the case of muon collider, we show that  the new s-channel and t-channel diagrams can increase the cross sections by virtue of their Yukawa couplings. We systematically compare our results for the muon collider with those obtained at the International Linear (ILC) and Compact Linear (CLIC) colliders. We select benchmark points and conduct signal-background analyses, incorporating detector simulations. For a 3 TeV muon collider, our results show an exclusion region at the 2$\sigma$ level and a discovery region at the 5$\sigma$ level.
\end{abstract}

\maketitle

\section{Introduction}

The discovery of the Higgs boson within the framework of the Standard Model (SM) by  ATLAS \cite{ATLAS:2012yve} and CMS \cite{CMS:2012qbp} marks the completion of the particle spectrum as defined by elementary particle physics. Nevertheless, several enigmas remain unresolved. These include both theoretical inquiries, such as the mechanism that stabilizes the electroweak scale and the generation of neutrino masses, as well as empirical observations, including the characteristics of particle dark matter and the asymmetry between matter and antimatter. 
Consequently, there exists a compelling rationale to explore theories beyond the Standard Model (BSM) within the proximity of the TeV scale \cite{Ait-Ouazghour:2020slc,Grzadkowski:2011jks,karahan2014effects, darvishi2018implication,Ouazghour:2018mld}. Numerous theories beyond the SM naturally incorporate an extended Higgs sector. A prevalent instance of this is the two-Higgs-doublet model (2HDM), which introduces an additional electroweak Higgs doublet, thereby enriching the theory with a diverse array of new Higgs boson phenomena and offering implications for flavor physics. Vigorous searches for these novel Higgs bosons have been underway at colliders, with the Large Hadron Collider (LHC) in particular playing a prominent role (for a comprehensive overview, we refer the reader to Ref. \cite{Kling:2020hmi} and the references cited therein). The absence of signal observation leads to the current bounds on the mass and couplings of those non-SM Higgs bosons. The High Luminosity LHC (HL-LHC) is poised to improve several of the earlier measurements and potentially find indications of new physics. However, to further improve the precise Higgs measurement program initiated at the LHC, it is mandatory to establish a controlled environment such as an electron-positron Higgs factory. 
This would allow for a comprehensive exploration of the characteristics of the recently discovered Higgs boson, akin to the Standard Model, and even facilitate the potential discovery of novel particles. Multiple projects to build $e^+e^-$ machines are currently in the planning stages. These include initiatives like the Circular Electron Positron Collider (CEPC) \cite{An:2018dwb}, the Compact Linear Collider (CLIC) \cite{CLICPhysicsWorkingGroup:2004qvu,Aicheler:2012bya}, the Future Circular Collider (FCC-ee) \cite{FCC:2018evy,TLEPDesignStudyWorkingGroup:2013myl}, and the International Linear Collider (ILC) \cite{LCCPhysicsWorkingGroup:2019fvj,Moortgat-Pick:2015lbx}. These projects aim to provide an ideal setting for intricate investigations into the properties of the Higgs boson, and potentially discover new particles. Recent investigations have indeed highlighted the remarkable prospects these colliders present in exploring the electroweak sector. This includes endeavors such as precise measurements of Higgs boson couplings \cite{Han:2020pif}, the detection of electroweak dark matter \cite{Han:2020uak,Belfkir:2023vpo}, study of neutrino \cite{Jueid:2023qcf,Jana:2023ogd} and the potential discovery of other BSM heavy particles \cite{Costantini:2020stv,Bandyopadhyay:2024plc,Han:2021udl}.

Charged Higgs bosons can be produced in multiple channels in hadron colliders. We refer to Ref. \cite{Akeroyd:2016ymd} for an extensive review of charged Higgs phenomenology. A light charged Higgs can be copiously produced from $t\bar{t}$ production followed by $t\to b H^+$ if kinematically allowed. The QCD processes $gb\to t H^-$ and  $gg \to  t\bar{b}  H^-$ \cite{Barger:1993th} are additional significant modes for singly-charged Higgs production. At the LHC, the production of a charged Higgs boson via $pp \to H^+ bj$ had been initially studied in the context of the Minimal Supersymmetric Standard Model (MSSM) \cite{Moretti:1996ra}.  In the 2HDM with various Yukawa textures, it was shown that  $pp \to H^+ bj$ is  sensitive to  $\tan \beta$  \cite{Arhrib:2015gra}.\\
Charged Higgs production  can proceed at ILC \cite{ILC:2013jhg,ECFADESYLCPhysicsWorkingGroup:2001igx,Bhattacharya:2023mjr} or 
CLIC \cite{CLICPhysicsWorkingGroup:2004qvu} through $e^+e^-\to \gamma^*, Z^* \to H^+H^-$ or $e^+ e^- \to \tau^{+} \nu_{\tau }H^{-}$ and $e^+ e^- \to t \bar{b} H^{-}$. At high energy, one can have also a vector boson fusion production:   $e^+ e^- \to \nu \bar{\nu}H^{+}H^-$\cite{Ouazghour:2023plc}. The production of a charged Higgs boson at muon collider is rather similar to $e^+e^-$, it can proceed through several processes. In Ref.\cite{Ouazghour:2023plc}, we studied the following ones:  i) pair production: $\mu^+ \mu^- \to H^+H^-$, ii) associated production with a gauge boson $\mu^+ \mu^- \to H^{\pm}W^{\mp}$, iii)  vector boson fusion $\mu^+ \mu^- \to \nu \bar{\nu} H^+ H^-$. 
 
Light charged Higgs bosons have been searched for in the past both at LEP \cite{DELPHI:2003eid}  and at Tevatron \cite{D0:2009hbc,D0:2009hbc} through the fermionic decays $H^+ \to\{ \tau^+ \nu, c\bar{s} , c\bar{b}\} $ channels. At the LHC, light charged Higgs bosons were investigated at  Run-1 in the decay channels $\tau^+ \nu$ \cite{ ATLAS:2014otc,CMS:2015lsf}, $c\bar{s}$ \cite{ATLAS:2013uxj,CMS:2015yvc}, and $c\bar{b}$ \cite{CMS:2016qoa}. There was no excess noted, and model independent limits are set on $BR(t \to H^+ b) \times BR(H^+ \to \tau^+ \nu)$. For charged Higgs mass heavier than 200 GeV, the decay modes $\tau^+\nu$ \cite{ATLAS:2016avi,CMS:2016szv} and $t\bar{b}$ \cite{ATLAS:2016qiq} are primarily investigated at Run-2 and set limits on cross sections times branching ratio. In the 2HDM, It should be noted that once the exotic decay channels into a lighter neutral Higgs, $H^\pm \to h W^\pm$ or $H^\pm \to A W^\pm$, are open \cite{Arhrib:2016wpw}, the existing ATLAS and CMS constraints are drastically lowered.
  
In this study, we investigate the production of charged Higgs bosons at a future muon collider in association with fermions: $\mu^+ \mu^- \to \tau^{+} \nu_{\tau }H^{-}$ and $\mu^+ \mu^- \to t \bar{b} H^{-}$ or in association with the W boson and a neutral Higgs bosons: $\mu^+ \mu^- \to W^\pm H^{\mp} S$, $S=h,H,A$, and $\mu^+ \mu^- \to W^\pm H^\mp Z$, within the framework of the 2HDM with varying Yukawa textures. \\ 
Regarding the production of a charged Higgs boson via $\mu^+ \mu^- \to \{ t \bar{b} H^{-},  \tau^{+} \nu_{\tau }H^{-} \} $ or $\mu^+ \mu^- \to W^\pm H^{\mp} S$  with $S=h,H,A$, it is noteworthy that multiple contributions come into play. These include pair production $\mu^+ \mu^- \to H^+H^-$, followed by the decay of one of the charged Higgs bosons into $H^+ \to \{ \tau^+ \nu,  t\bar{b}\}$, $H^+\to \{ W^+ h,\, W^+ H,\, W^+A \}$. Furthermore, the $t \bar{t}$ pair production process $\mu^+ \mu^- \to t \bar{t}$ contributes as well to $\mu^+ \mu^- \to t \bar{b} H^{-}$ if one of the top quarks decay into $H^+$ and $b$, particularly when the charged Higgs boson mass $(m_{H^{\pm}})$ is less than that of the top quark $(m_{\text{top}})$. In the case of the associate production with neutral Higgs boson, we can also have resonant production of the charged Higgs coming from the on-shell production $e^+e^-\to h\,A$ or $e^+e^- \to H\,A$ followed by the decay $A \to W^\pm H^\mp$ or $H\to W^\pm H^\mp$.

   
 In our study, we present numerical results for different Yukawa textures of the 2HDM corresponding to cross sections at a muon collider. Of particular interests are 2HDM types II and X, where the neutral Higgs coupling to a pair of muons as well as $ t \bar{b} H^-$ and $\tau^+ \nu H^-$ couplings receive large $\tan\beta$ enhancement and could potentially increase the cross section of $\mu^+ \mu^- \to \{t\bar{b} H^-, \tau^+ \nu H^-\} $ and $e^+ e^- \to \{ t \bar{b} H^-, \tau^+ \nu H^-\}$. We will also present our findings for $ \ell^+\ell^- \to W^\pm H^{\mp} S$, $S=h,H,A$ at the future $e^+e^-$ and muon colliders. Our numerical outcomes are provided after thoroughly exploring the 2HDM parameter space, adhering to various theoretical (perturbative unitarity, perturbativity, and vacuum stability) and experimental (derived from SM-like Higgs boson discovery data, BSM Higgs boson exclusion data, Electroweak precision tests (EWPT), and flavor physics) constraints. We perform signal and various SM background calculations, conduct a comprehensive Monte Carlo (MC) analysis, and gauge the sensitivity at a center-of-mass energy of 3 TeV.

The structure of the paper is as follows: we start with a concise overview of the 2HDM in the subsequent section, covering the scalar sector, necessary couplings, and pertinent theoretical and experimental constraints. Moving forward, in Section \ref{section3}, we give tree level Feynman diagrams for $\mu^+ \mu^- \to \tau^+ \nu H^- $, 
$\mu^+ \mu^- \to t \bar{b} H^-$,  $\mu^+ \mu^- \to W^\pm H^{\mp} S$, $S=h,H,A$, and $\mu^+ \mu^- \to W^\pm H^\mp Z$. Section \ref{Section4} then presents the numerical outcomes of our investigation, considering theoretical and flavor physics constraints as well as a set of experimental constraints from LEP-II, Tevatron, and LHC. In Section \ref{section5}, we outline the specifics of the Monte Carlo analysis and establish the significance of the potential discovery of charged Higgs at the 3 TeV muon collider. Finally, our concluding remarks are presented in Section \ref{sec:conclusion}.

\section{review on the 2HDM}
\label{sec:model}
\subsection{Model review}
\label{subsec:review}
In this section, we briefly discuss the basic features of the 2HDM \cite{Lee:1973iz,Branco:2011iw} and the various Yukawa textures \cite{Paschos:1976ay,Glashow:1976nt}. Hence, in the 2HDM, in addition to the SM doublet $\Phi_1$, a new doublet $\Phi_2$ with hypercharge $+1$ is added to the Higgs sector, assuming that CP is not spontaneously broken. The two Higgs scalar doublets can be parametrized as follows :
\begin{equation}
\Phi_1 = \left(
\begin{array}{c}
\phi_1^+ \\
\phi_1^0 \\
\end{array}
\right)
\quad {\rm and}\quad 
\Phi_2 = \left(
\begin{array}{c}
\phi_2^+ \\
\phi_2^0 \\
\end{array}
\right)
\end{equation}
with $\phi_1^0 = (v_1+\psi_1+ i \eta_1)/\sqrt{2}$, $\phi_2^0 = (v_2+\psi_2+ i \eta_2)/\sqrt{2}$ and $\sqrt{v_1^2+v_2^2}=v=246$ GeV. The general scalar potential invariant under the electroweak gauge group $SU(2)_L\times U(1)_Y$ can be expressed as \cite{Branco:2011iw}:
\begin{eqnarray}
V(\Phi_1,\Phi_2) &=& m_{11}^2 \Phi_1^\dagger\Phi_1+m_{22}^2\Phi_2^\dagger\Phi_2-[m_{12}^2\Phi_1^\dagger\Phi_2+{\rm h.c.}] + \frac{\lambda_1}{2}(\Phi_1^\dagger\Phi_1)^2 + \frac{\lambda_2}{2}(\Phi_2^\dagger\Phi_2)^2\nonumber\\
&+&\lambda_3(\Phi_1^\dagger\Phi_1)(\Phi_2^\dagger\Phi_2)
+\lambda_4(\Phi_1^\dagger\Phi_2)(\Phi_2^\dagger\Phi_1) 
+\left\{\frac{\lambda_5}{2}(\Phi_1^\dagger\Phi_2)^2+h.c\right\}  \label{pot1}
\label{scalar_pot}
\end{eqnarray}
In the above potential, all the $m_{11}^2$, $m_{22}^2$, and $m_{12}^2$ parameters as well as the $\lambda_{i}\,(i=1,2,3,4,5)$ couplings are assumed to be real to ensure that our potential is CP-conserving. We also advocate a discrete $Z_2$ symmetry in order to avoid Flavor Changing Neutral Current (FCNC) at the tree level once applied to the fermionic sector. Such a $Z_2$ symmetry is only softly broken by the bilinear term proportional to $m_{12}^2$ parameter. 

After electroweak symmetry breaking, the 8 degrees of freedom initially present in the two Higgs doublet fields are reduced. Three of these degrees of freedom are absorbed by the Goldstone bosons, giving mass to the gauge bosons $W^\pm$ and Z, and we are left with five physical Higgs states : a pair of charged Higgs $H^\pm$, a CP-odd state $A$ and 2 CP-even states : $H$ and $h$ with $m_h < m_H$. One of the neutral CP-even Higgs would be identified as the 125 GeV Higgs-like particle observed at the LHC. 
The combination $v^2=v_1^2+v_2^2=(2\sqrt{2} G_F)^{-1}$ can be used to fix one of the vacuum expectation value (vev) as a function of $G_F$ and $\tan\beta$. Together with the two minimization conditions, the scalar potential in eq.(\ref{scalar_pot}) has seven independent parameters:
\begin{equation}
\label{eq:modelpara}
\alpha,\quad \tan\beta=\frac{v_2}{v_1},\quad  m_{h}\,(\equiv m_{125}),\quad m_{H},\quad m_A,\quad 
m_{H^\pm}\quad \ \rm{and}\ \  m_{12}^2,
\end{equation}
where $\alpha$ and  $\beta$ are respectively  the CP-even and CP-odd mixing angles. 
In this work, we assume that $h$ is  the observed SM-like boson at the LHC with $m_h=125$ GeV, so the scalar potential described only by six free parameters. 

In the Yukawa sector, assuming that both Higgs doublets couple to all fermions, like in the SM, we will end up with a large tree level FCNCs mediated by the neutral Higgs scalars. To prevent such large FCNCs at the tree level, the 2HDM needs to satisfy Paschos-Glashow-Weinberg theorem \cite{Paschos:1976ay,Glashow:1976nt} which asserts that all fermions with the same quantum numbers couple to the same Higgs multiplet. One can have 4 different types of Yukawa textures depending on how the doublets $\Phi_1$ and $\Phi_2$ interact with the fermions.
In the 2HDM type-II, the type used in the MSSM, $\Phi_2$ interacts with up-type quarks and $\Phi_1$ interacts with the charged leptons and down-type quarks while in the 2HDM Type X, $\Phi_2$ couples to all quarks while $\Phi_1$ couples to all leptons.
In terms of the mass eigenstates of the neutral and charged Higgs bosons, the Yukawa interactions can be written as:
 \begin{eqnarray}
-{\cal L}_Y &=& \sum_{f=u,d,\ell} \frac{m_f }{v} \left[ \kappa^f_h  \bar f f  h + \kappa^f_H \bar f f H - i \kappa^f_A \bar f \gamma_5 f A \right]  +  \label{eq:Yukawa_CH} \\
&&+ \frac{\sqrt{2}}{v} \left[ \bar u_{i} V_{ij}\left( m_{u_i}  \kappa^{u}_A P_L + \kappa^{d}_A  m_{d_j} P_R \right)d_{j}  H^+ \right]   
+ \frac{\sqrt{2}}{v}  \bar \nu_L  \kappa^\ell_A m_\ell \ell_R H^+ +\hc \nonumber 
\end{eqnarray} 

\begin{table}[!h]
\begin{center}
\begin{tabular}{|c|c|c|c|c|c|c|c|c|c|}
 \hline 
 & $\kappa_h^u$ & $\kappa_h^d$ & $\kappa_h^l$ & $\kappa_H^u$ & $\kappa_H^d$ & $\kappa_H^l$ & $\kappa_A^u$ & $\kappa_A^d$ & $\kappa_A^l$ \\
\hline 
Type-X & $c_\alpha/s_\beta$ & $c_\alpha/s_\beta$& $-s_\alpha/c_\beta$ & $s_\alpha/s_\beta$ & $s_\alpha/s_\beta$ & $c_\alpha/c_\beta$ & $c_\beta/s_\beta$ & 
$-c_\beta/s_\beta$ & $\tan\beta$ \\ \hline
Type-II & $c_\alpha/s_\beta$ & -$s_\alpha/c_\beta$& $-s_\alpha/c_\beta$ & $s_\alpha/s_\beta$ & $c_\alpha/c_\beta$ & $c_\alpha/c_\beta$ & $c_\beta/s_\beta$ & 
$\tan\beta$ & $\tan\beta$ \\ \hline
\end{tabular}
\end{center}
\caption{Yukawa couplings of the $h$, $H$, and $A$ Higgs bosons to the quarks and leptons in  2HDM Type II and X.} 
\label{coupIII}
\end{table} 

Whilst the reduced coupling of the lighter Higgs boson, $h$, to either $WW$ or $ZZ$ is given by $\sin(\beta-\alpha)$, on the other hand, the coupling of the heavier Higgs boson, $H$, is equivalent to the SM coupling multiplied by $\cos(\beta-\alpha)$. Notably, the coupling between the pseudoscalar, $A$, and vector bosons is absent due to CP symmetry invariance. 

Since these couplings are crucial throughout this study, it would indeed be better to explicitly note the following identities,
\begin{eqnarray}
\frac{\cos\alpha}{\sin\beta} & = & s_{\beta-\alpha}+\cot\beta\,c_{\beta-\alpha} \nonumber\\
-\frac{\sin\alpha}{\cos\beta} & = & s_{\beta-\alpha}-\tan\beta\,c_{\beta-\alpha} \nonumber \\
\frac{\sin\alpha}{\sin\beta} & = & c_{\beta-\alpha}-\cot\beta\,s_{\beta-\alpha}  \nonumber\\
\frac{\cos\alpha}{\cos\beta} & = & c_{\beta-\alpha}+\tan\beta\,s_{\beta-\alpha} \label{eq:trigonometricexp}
\end{eqnarray}
It is clear that $-\frac{\sin\alpha}{\cos\beta}$ and $\frac{\cos\alpha}{\cos\beta}  $ exhibit some enhancement for large $\tan\beta$. Note that close to the decoupling limit $\sin(\beta-\alpha)\approx 1$, which is also favored by LHC data, $h$ coupling to fermions reduces to unity.

For completeness, we also list the Feynman rules for pure scalar interactions :
\begin{eqnarray}
&&g_{h H^+H^-} =  -\frac{1}{v}\bigg[ \Big( 2m_{H^\pm}^2-m_h^2 \Big) s_{\beta-\alpha} + \Big(m_h^2 - 2\frac{m_{12}^2}{s_{2\beta}} \Big)\frac{c_{\beta+\alpha}}{s_\beta c_\beta} \bigg] \nonumber\\
&& g_{H H^+H^-}  =  -\frac{1}{v}\bigg[ \Big( 2m_{H^\pm}^2-m_H^2 \Big) c_{\beta-\alpha} +\Big(m_H^2 - 2\frac{m_{12}^2}{s_{2\beta}} \Big)\frac{s_{\beta+\alpha}}{s_\beta c_\beta} \bigg] \nonumber\\
&&g_{Hhh} = -\frac{c_{\beta-\alpha}}{v s_{2\beta}^2}\bigg[  (2 m_h^2 + m_H^2) s_{2\alpha} s_{2\beta}   - 2 m_{12}^2 
(3  s_{2\alpha}  - s_{2\beta} )\bigg] \nonumber\\
&&g_{A H^+G^-} =  \frac{1}{v}(-m_A^2 + m_{H^\pm}^2) \nonumber\\
&&g_{h H^+G^-} =  -\frac{c_{\beta-\alpha}}{v}(m_h^2 - m_{H^\pm}^2) \nonumber\\
&&g_{H H^+G^-}=   -\frac{s_{\beta-\alpha}}{v}(m_H^2 - m_{H^\pm}^2)\label{thdm:h1hphm}
\end{eqnarray}

The relevant part of the Lagrangian describing the interactions of the gauge bosons with scalars is :
\begin{eqnarray}
{\cal L} &=& \frac{g}{2}W_{\mu}^+ ( (H^- \stackrel{\leftrightarrow}{\partial}^{\mu} A)-
 ic_{\beta-\alpha} (H^- \stackrel{\leftrightarrow}{\partial}^{\mu} h)+
  i s_{\beta-\alpha} (H^- \stackrel{\leftrightarrow}{\partial}^{\mu} H))+ h.c\nonumber\\
&&+ ( ie A_\mu+ i\frac{g (c_W^2-s_W^2)}{2 c_W} Z_{\mu}^+)  (H^\mp\stackrel{\leftrightarrow}{\partial}^{\mu} H^\pm )  + 
 i\frac{g}{2 c_W} Z_{\mu}  (-s_{\beta-\alpha} H\stackrel{\leftrightarrow}{\partial}^{\mu} A +  c_{\beta-\alpha} h\stackrel{\leftrightarrow}{\partial}^{\mu} A)  +\nonumber\\
 &&  \frac{igm_W}{c_W^2}g^{\mu \nu}  (W_\mu W_\nu +  c_W^2 Z_\mu Z_\nu )
(  s_{\beta-\alpha} h + c_{\beta-\alpha}H) + \\
&& \frac{-ie^2}{2 s_W} W_\mu^\pm A_\nu  H^\mp ( -c_{\beta-\alpha} h + s_{\beta-\alpha}H + A) g^{\mu \nu} +
\frac{ie^2}{2 c_W}  W_\mu^\pm Z_\nu  H^\mp ( -c_{\beta-\alpha} h + s_{\beta-\alpha} H + A) g^{\mu \nu}  \nonumber \label{lag}
\end{eqnarray}

\subsection{Theoretical and Experimental Constraints}
\label{constraint}
The phenomenological analysis in 2HDM is conducted by imposing a comprehensive set of theoretical constraints as well as exclusion limits from experimental measurements at colliders such as LEP-II, Tevatron, and LHC. These are:

\begin{itemize}
\item \textbf{Unitarity}: to ensure perturbative unitarity in scattering processes, we adopt constraints from Refs. ~\cite{Kanemura:1993hm,Akeroyd:2000wc,Arhrib:2000is,Kanemura:2015ska}.
\item \textbf{Perturbativity}:  to confine quartic couplings within $|\lambda_i| < 8 \pi$ for each $i = 1, \ldots, 5$~\cite{Branco:2011iw}.
\item \textbf{Vacuum stability}: that guarantees positivity in all field directions $\Phi_1$, $\Phi_2$ to prevent boundedness from below (BFB), and as a consequence the following conditions \cite{Barroso:2013awa,deshpande1978pattern}, 
\begin{eqnarray}
\lambda_{1,2}>0,  \quad
\lambda_3>- \sqrt{\lambda_1\lambda_2}, \quad
\lambda_3+\lambda_4-|\lambda_5|> - \sqrt{\lambda_1\lambda_2},
\label{eq:VB}
\end{eqnarray}
must be satisfied in the whole parameter space.
\item[\textbullet] \textbf{Electroweak precision observables}: 
It is common that in order to  constrain BSM physics we use the global electroweak fit through the oblique parameters $S$, $T$ and $U$ \cite{Peskin:1991sw}. In the SM, it is well known that EWPT imply a close relationship between the three masses, $m_t$, $m_h$, and $m_W$. Similarly, in the 2HDM, EWPT imply constraints on the splitting between the extra Higgs bosons. From the Particles Data Group (PDG) review \cite{Lu:2022bgw}, with a fixed $U = 0$, the best fit of $S$, $T$ parameters is given by:
\begin{eqnarray}
S &=& 0.05 \pm 0.08, \quad T = 0.09 \pm 0.07, \quad \rho_{ST} = 0.92 \nonumber
\end{eqnarray}
where $\rho_{ST}$ is the correlation parameter. The analytical expressions for $S$ and $T$ in 2HDM are taken from \cite{Kanemura:2015mxa}.			
\end{itemize}
The constraints mentioned above have been incorporated into \texttt{2HDMC-1.8.0} \cite{Eriksson:2009ws}, a publicly available code. 
This tool is utilized to systematically explore the parameter space of the 2HDM, ensuring its compatibility with the specified constraints, 
and to calculate the Higgs branching ratios at each point. 
Moreover, \texttt{2HDMC} features interfaces to both \texttt{HiggsBounds-5.10.1} \cite{Bechtle:2020pkv, Bechtle:2015pma} and 
\texttt{HiggsSignals-2.6.1} \cite{Bechtle:2020uwn}. These interfaces facilitate further analysis by incorporating experimental 
exclusion limits and signal strength measurements, respectively, providing a comprehensive framework for testing the viability of 
2HDM parameter sets against LEP-II, Tevatron, and the current LHC data.
Moreover, the parameter space is examined using flavor constraints as well. To address the B-physics observables, we used the \textbf{SuperIso v4.1} tool \cite{Mahmoudi:2008tp} to calculate the relevant observables. Consistency checks were then performed at a 2$\sigma$ confidence level, taking into account the available experimental measurements as reported in \cite{Mahmoudi:2008tp, HFLAV:2016hnz, Haller:2018nnx}, displayed in Table \ref{Tab2}.
%
\begin{table}[!hb]
\centering
\setlength{\tabcolsep}{7pt}
\renewcommand{\arraystretch}{1.2} %
\begin{tabular}{|l||c|c|}
\hline
Observable&Experimental result&95\% C.L. Bounds\\\hline
BR($B_{u}\to \tau\nu$)\cite{Haller:2018nnx}&$(1.06 \pm 0.19) \times 10^{-4}$&$ [0.68\times 10^{-4} , 1.44\times 10^{-4} ]$\\\hline
BR($B_{s}^{0}\to \mu^{+}\mu^{-}$)\cite{Haller:2018nnx}&$(2.8 \pm 0.7) \times 10^{-9}$&$[1.4 \times 10^{-9}, 4.2\times 10^{-9}]$\\\hline
BR($B_{d}^{0}\to \mu^{+}\mu^{-}$)\cite{Mahmoudi:2008tp}&$(3.9\pm 1.5)\times10^{-10}$&$[0.9\times 10^{-10}, 6.9\times10^{-9}$\\\hline
BR($\bar{B}\to X_{s}\gamma$)\cite{Haller:2018nnx,HFLAV:2016hnz}&$(3.32\pm 0.15)\times10^{-4}$&$[3.02\times 10^{-4} , 3.61\times 10^{-4}]$\\\hline
\end{tabular}
\caption{Experimental results of $B_{u}\to \tau^+\nu$, $B_{s,d}^{0}\to \mu^{+}\mu^{-}$ and $\bar{B}\to X_{s}\gamma$ at 95$\%$ C.L.}
\label{Tab2}
\end{table}

\section{Computational Procedure steps}
\label{section3}
In this section, we list every process that is investigated in this study, namely 
$\ell^-\ell^+ \to \tau^+\nu_{\tau} H^{-}$, $\ell^-\ell^+ \to t\bar{b} H^{-}$,  $\ell^-\ell^+ \to H^\pm W^\mp Z$ and $\ell^-\ell^+ \to S H^\pm W^\mp$ $(S=h, H, A)$  where $\ell$ could be muon or electron. 
 \subsection{$\ell^-\ell^+ \to \tau^+\nu_{\tau} H^{-}$ and $\ell^-\ell^+ \to t\bar{b} H^{-}$}
\begin{figure}[!h]
\centering
\includegraphics[width=0.8\textwidth]{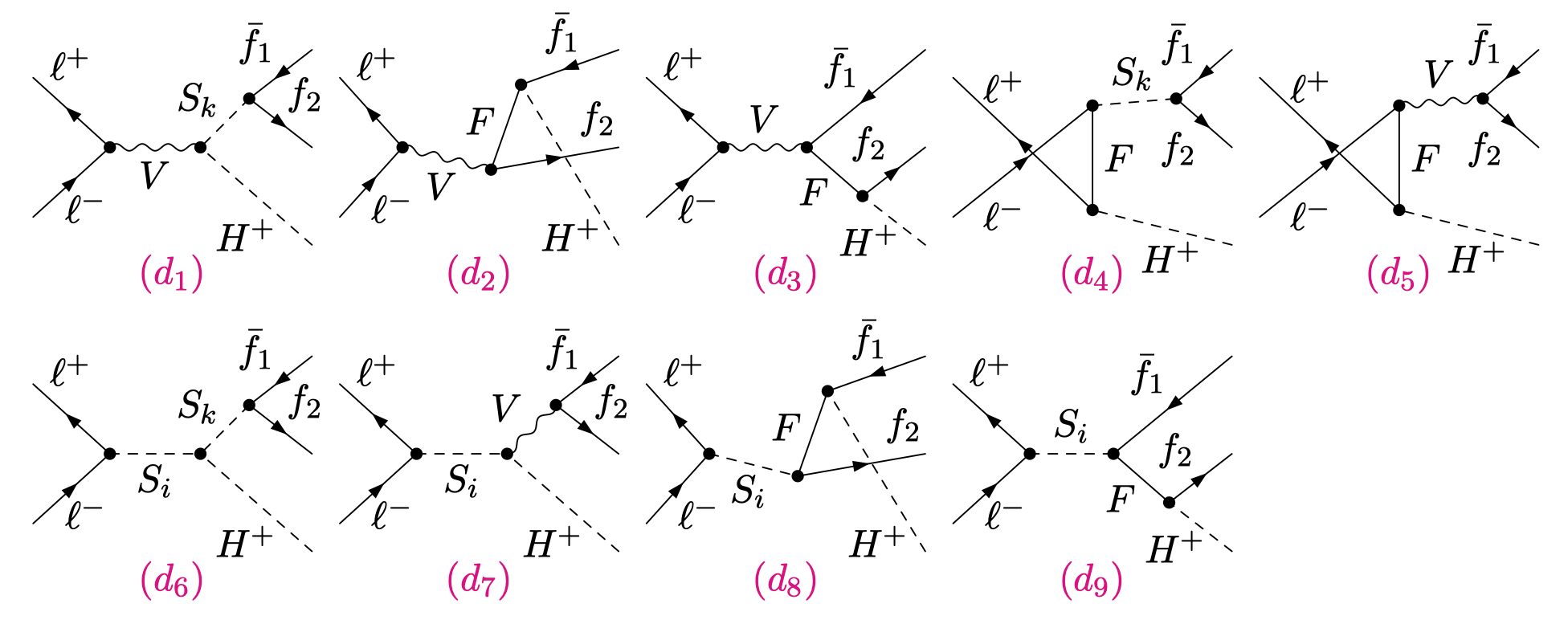}
\caption{Tree level generic Feynman diagrams for $\ell^-\ell^+ \to f_2\bar{f_1} H^{+}$ where $\ell = e,\,\mu$ and $(\bar{f_1},\,f_2) = (\tau^+,\,\nu_\tau)$ or $(\bar{t},\,b)$ at muon collider in the 2HDM. The $S_i$, $S_k$, $V$ and $F$ propagators stand for ($h, H, A$), ($H^\pm, G^\pm$), ($\gamma,Z,W^\pm$) and ($\tau,\nu_\ell$) respectively. We note here that diagram $(d_8)$ is absent for the $\tau^-\bar{\nu_{\tau}}H^+$ production.}
\label{fig:fig1}
\end{figure}

The Feynman diagrams, which contribute at tree level to both processes, are depicted in Fig.\ref{fig:fig1}. 
We utilize the public Mathematica packages \texttt{FeynArts} \cite{Hahn:2000kx} and \texttt{FormCalc}\cite{Hahn:1998yk} to generate the amplitudes and to compute the cross sections at a given center of mass-energy. The calculation is performed in the 't Hooft-Feynman gauge.

The phenomenology of charged Higgs bosons can be investigated using the kinematically permitted production processes, either through $2\to2$ processes like $\mu^+ \mu^- \to  H^+ H^-$ and $\mu^+ \mu^- \to W^\pm H^\mp $ or via $2\to3$ processes such as $\mu^+ \mu^- \to \tau^+ \nu_{\tau} H^-$ and $\mu^+ \mu^- \to t\bar{b} H^- $. The first two processes are discussed in detail in Ref. \cite{Ouazghour:2023plc}. The last two processes are generated by the s-channel diagrams: by the $Z$ gauge boson and photon exchange, as shown in Fig.\ref{fig:fig1}-$(d_{1})$, $(d_{2})$, and $(d_{3})$, and by the neutral Higgs $h$, $H$, and $A$ exchange, as can be seen from $(d_{6})$, $(d_{7})$, and $(d_{8})$ diagrams, as well as by the t-channel diagram with neutrino exchange, as illustrated by diagrams $(d_{4})$ and $(d_{5})$ in Fig.\ref{fig:fig1}. It is important to note that the CP-even Higgs ($H$) or the CP-odd ($A$) could both mediate resonant production if the center of mass energy is close to the sum of the masses of the particles in the final state. In the case of a muon collider with 3 TeV or 10 TeV center of mass energy, there is no resonant effect to look for since we are interested only in Higgs masses less than 1 TeV. To stabilize numerical integration in the $2\to 3$ phase space, we have included the partial widths for all internal exchanges of Higgs and gauge bosons. The $2 \to 2$ processes  $\mu^+ \mu^- \to  H^+ H^-$ (resp $\mu^+ \mu^- \to  W^{+*} H^-$) followed by the decay $H^+ \to \tau^+ \nu $ (resp. $W^{+*} \to \tau \nu $), can both give rise to the current $2 \to 3$ processes. In the case of light charged Higgs production, we have verified that the cross section of  $\mu^+ \mu^- \to \tau^+ \nu_{\tau} H^-$ (respectively  $\mu^+ \mu^- \to t \bar{b} H^-$)  can be obtained from $\sigma(\mu^+ \mu^- \to  H^+ H^-)\times Br(H^+\to \tau^+ \nu)$ (respectively $\sigma(\mu^+ \mu^- \to  H^+ H^-)\times Br(H^+\to t\bar{b} ))$  as dictated by the narrow-width approximation and this provides a good check of our calculation. A careful examination of the Feynman diagrams reveals that Fig.\ref{fig:fig1}-$(d_{8})$ with $S_i=h,\,H,\,A$ and Fig.\ref{fig:fig1}-$(d_{4})$ with $S_k=H^+$ contain 3 vertices that are proportional to $\tan\beta$  in the 2HDM type II and X. This will make the square amplitude have a term scaling as $\tan^6\beta$ which could enhance the cross section. Note that in the case of $e^+e^-$, all Feynman diagrams with neutral Higgs exchange in the s-channel and diagrams with fermion exchanges in the t and u channels would be proportional to the electron mass and are therefore negligible. In such a case we keep only diagrams with photon and Z exchange in the s-channel.
 
 \begin{figure}[!hb]
\centering
\includegraphics[width=0.8\textwidth]{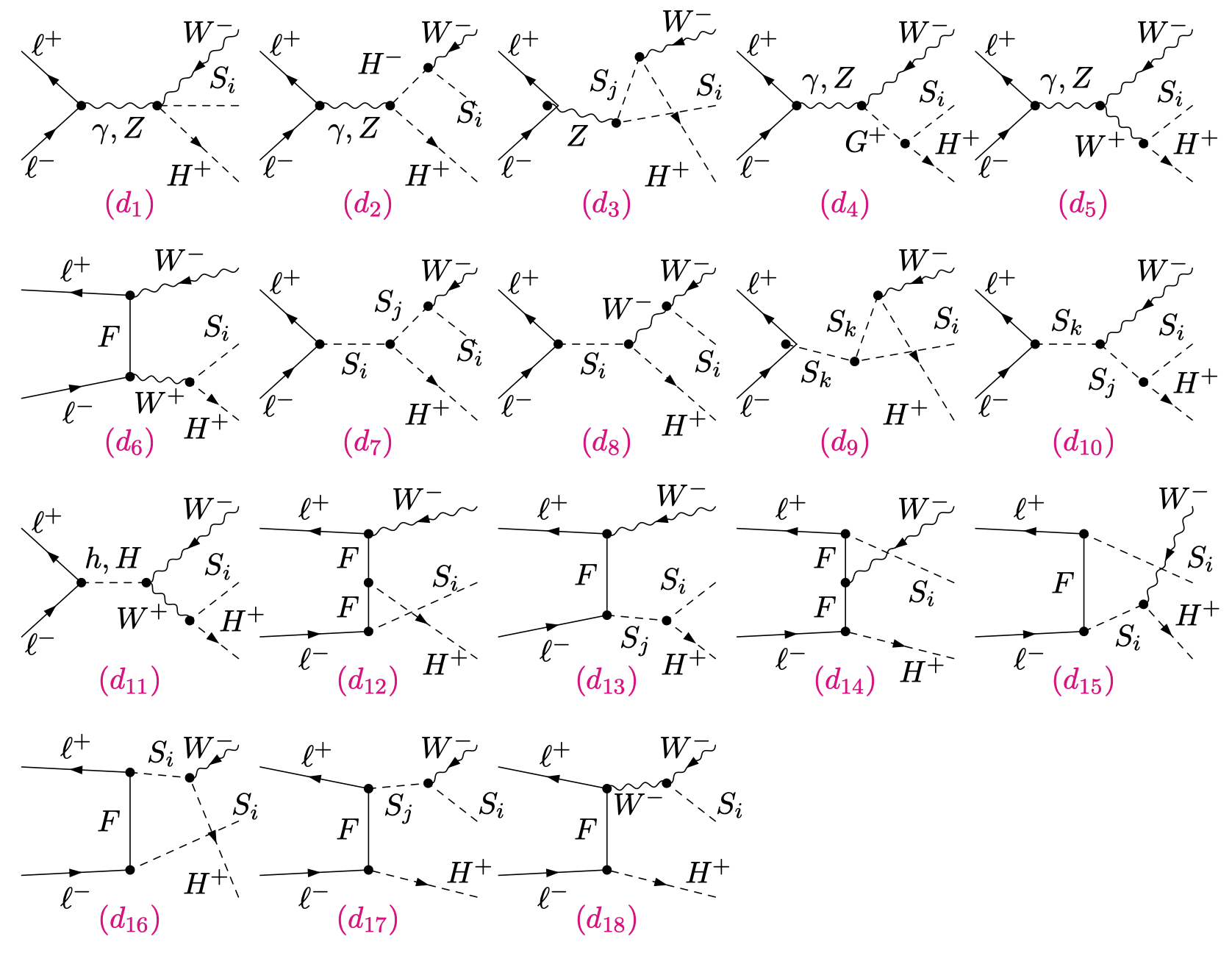}
\caption{Tree level generic Feynman diagrams for $\ell^-\ell^+ \to H^{\pm} W^\mp S$ ($S=h,H,A$) are shown in $(d_{1,...,6})$. For all diagrams $S_i=h,H,A$.
For $d_3$, if $S_i=h$ or $H$, $S_j$ should be A, while if $S_i=A$, $S_j$ should be either $h$ or $H$. All diagrams $(d_{7,...,18})$
are proportional to lepton mass, these diagrams are taken into account for  $\mu^-\mu^+ \to H^{\pm} W^\mp S$.}
\label{fig:fig2}
\end{figure}

\begin{figure}[!hb]
\centering
\includegraphics[width=0.6\textwidth]{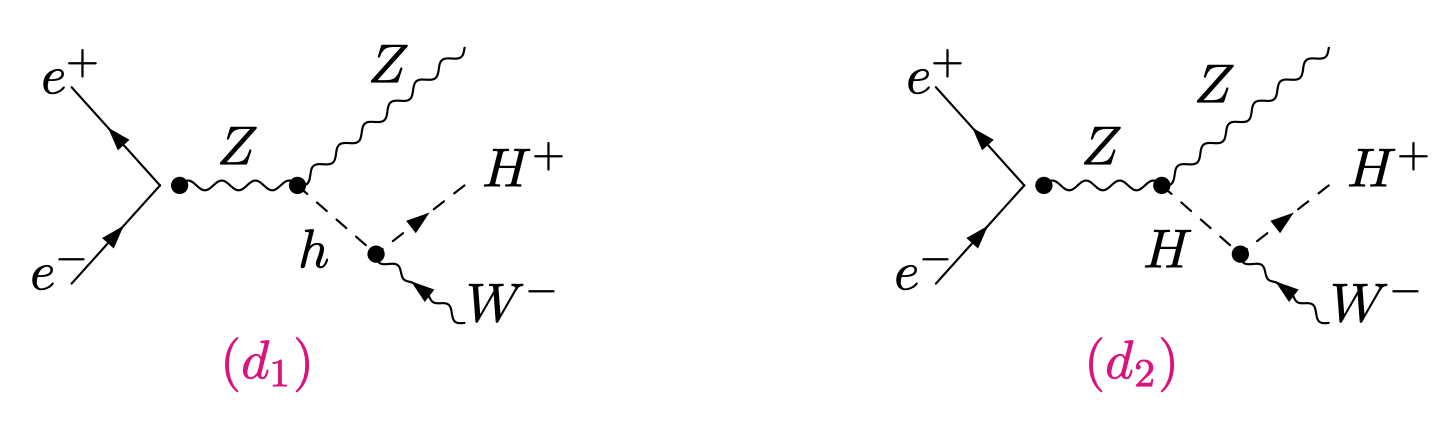}
\caption{Tree level Feynman diagrams for $e^+e^- \to H^{\pm} W^\mp Z$  in the 2HDM. }
\label{fig:fig3}
\end{figure}

\subsection{$l^+l^- \to H^\pm W^\mp Z$ and $l^+l^- \to H^\pm W^\mp S$ $(S=h, H, A)$}
The processes $e^+e^- \to H^\pm W^\mp Z$ and $e^+e^- \to H^\pm W^\mp S$ $(S=h, H, A)$ have been studied in Ref.\cite{Kanemura:2000cw} within the MSSM. In Fig.\ref{fig:fig2}, except for the final Feynman diagram $d_{6}$ that involves t-channel mediation by neutrino exchange, the majority of the diagrams for the $e^+e^- \to H^\pm W^\mp S$ process  are mediated by the s-channel photon or Z exchange. The couplings involved in $ e^+e^- \to H^\pm W^\mp S$ are pure gauge couplings. Only $H^\pm W^\mp S$, $W^\pm H^\mp A_\nu S$, and $ W^\pm H^\mp Z S$, with $S=h, H$, come either with the mixing factor $\cos(\beta-\alpha)$  or $\sin(\beta-\alpha)$. The Feynman diagrams that contribute to $\mu^+ \mu^- \to W^\pm H^\mp S$  are proportional either to the muon mass or the muon mass square are depicted in Fig.\ref{fig:fig2}-$(d_{7,...,18})$. In the case of type-X 2HDM, $d_{7,8,9,10,11,13,18}$ are proportional to $\tan\beta$ while $d_{12,14,15,16,17}$ are proportional to $\tan^2\beta$. Therefore the amplitude square for $\mu^+ \mu^- \to W^\pm H^\mp S$ will contain a term proportional to $\tan^4\beta$.

The process $e^+e^- \to H^\pm W^\mp Z$ drawn in Fig.\ref{fig:fig3} is mediated by s-channel Z exchange. The charged Higgs is radiated either from $h$ or $H$ :  $e^+e^- \to Zh^* \to Z H^\pm W^\mp $ or $e^+e^- \to ZH^* \to Z H^\pm W^\mp $. The couplings $ZZh$ and $hW^\pm H^\mp$ are respectively proportional to $\sin(\beta - \alpha)$ and $\cos(\beta - \alpha)$, whereas the couplings $ZZH$ and $HW^\pm H^\mp$ are respectively proportional to $\cos(\beta - \alpha)$ and $\sin(\beta - \alpha)$. Therefore the squared amplitude of  $e^+e^- \to H^\pm W^\mp Z$ is proportional to $\cos^2(\beta - \alpha) \sin^2(\beta - \alpha)$. Given that $\cos(\beta - \alpha)$ is constrained by the LHC data to be quite small, we anticipate this cross section to be rather suppressed.

\section{Numerical results}
\label{Section4}
We perform random scans over the parameter space of the 2HDM within the following ranges:
\begin{eqnarray}
\centering
&&m_{h} = 125.09\ \text{GeV}, \ \ m_{H} \in [130,~1000]\ \text{GeV},\hspace{0.3cm} \sin(\beta-\alpha)\in [0.97,1], \ \  \nonumber \\ && m_{A, H^\pm}\in [80,~1000]\ \text{GeV}, 
  \ \  \tan\beta\in [0.5,~45],\  \hspace{0.3cm}m_{12}^2 \in [0,10^6]\  \text{GeV$^2$};,
\label{parm}
\end{eqnarray}
where we have assumed that the lightest Higgs state $h$ is the observed SM-like Higgs boson at the LHC \cite{ATLAS:2012yve,CMS:2012qbp} and have set $m_h=125$ GeV. After scrutinizing the parameter space of the model with the
theoretical and experimental constraints described above, the resulting parameter space points  will be passed to
FormCalc \cite{Hahn:2001rv,Hahn:1998yk,Kublbeck:1990xc} to compute the corresponding cross section of each process at the lepton collider. It is worth to mention here that the corresponding cross section for each charge-conjugate 
process remains unchanged; and what is presented below is their combined sum.

Note that in the scan above, we consider the charged Higgs mass in the range 80-1000 GeV. This is because in 2HDM type X, one can still have a relatively light charged Higgs 80-160 GeV that is consistent with all B-physics constraints \cite{Enomoto:2015wbn} as well as with LEP-II, Tevatron, and LHC data, and this is mostly ascribed to the dominance of other channels such as $H^+\to W^+ A$ or $H^+ \to W^+ h$ \cite{Arhrib:2016wpw}, which suppresses charged Higgs searches via $H^+\to \tau^+ \nu$. Since the charged Higgs mass in the 2HDM-II is severely constrained by $Br(B\to X_s \gamma)$, we consider $m_{H^\pm}>800$ GeV for all $\tan\beta$ values in this analysis \cite{Misiak:2020vlo}. Also take note of the fact that large $\tan\beta$ would be ruled out in the 2HDM type II by the LHC's $H,A\to \tau^+ \tau^-$ and/or $H^+\to \tau^+ \nu$ searches.

Before presenting our numerical analysis, let's recall here that a wide phenomenological scenarios could occur depending on the charged Higgs boson mass, regardless of the center-of-mass energy. As an example, for $m_{H^+} \gtrsim170$ GeV ($m_{H^+} \lesssim 170$ GeV), the dominant decay mode is $H^+ \to t \bar{b}$ ($H^+ \to \tau^+ \nu_\tau$). The $H^+ \to W^+ A$, however, embraced the scene, with charged Higgs boson mass lying in $\sim\,[156,\,640]$ GeV; and a significant competition emerges between the decay modes $H^+ \to W^+h$ and $H^+\to W^+H$ as the charged Higgs mass increases within the range $[200\sim640]$ GeV. Finally, for $m_{H^+}>640$ GeV, only the decay mode $H^+ \to W^+H$ remains dominant.

\subsection{$e^+ e^- \to \tau^+\nu_\tau H^- ,\, t \bar{b} H^-$ processes}

\begin{figure}[!ht]
\centering
\includegraphics[width=0.3\textwidth]{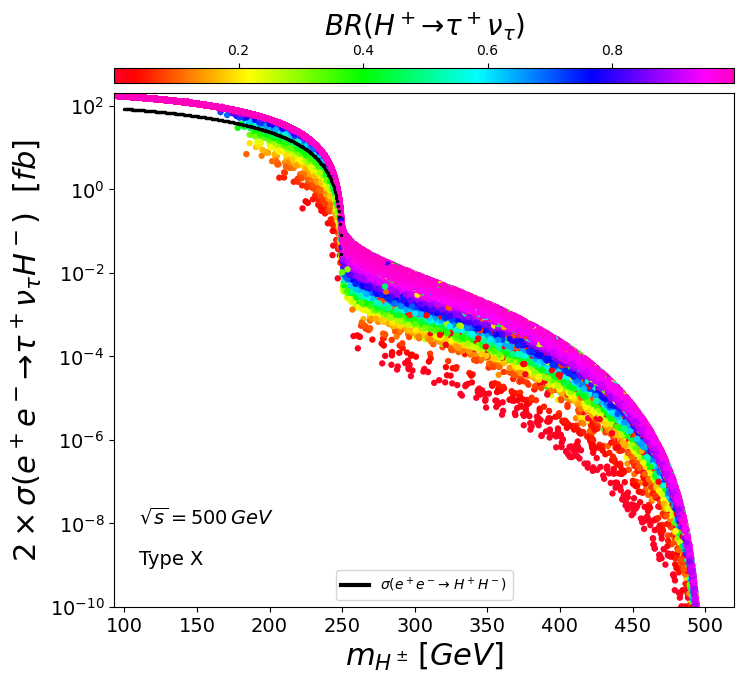}
\includegraphics[width=0.3\textwidth]{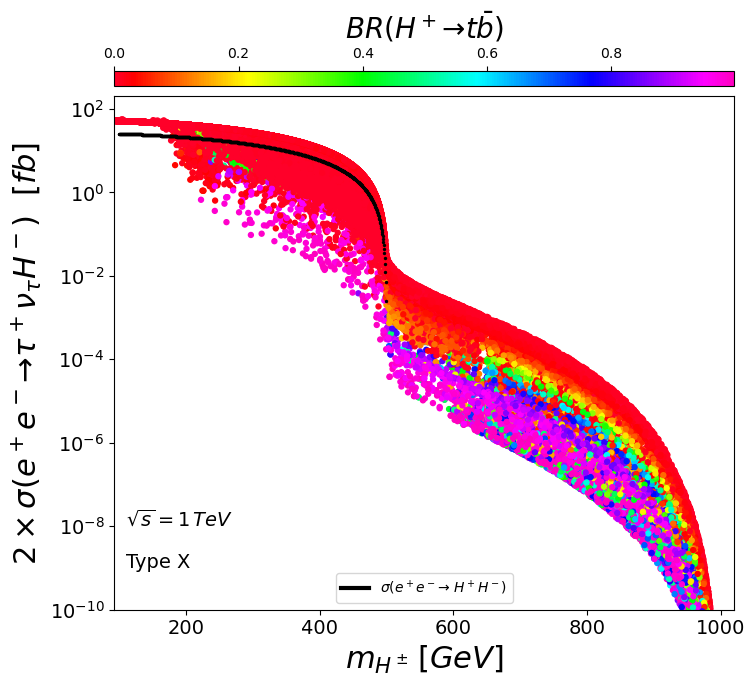}
\includegraphics[width=0.3\textwidth]{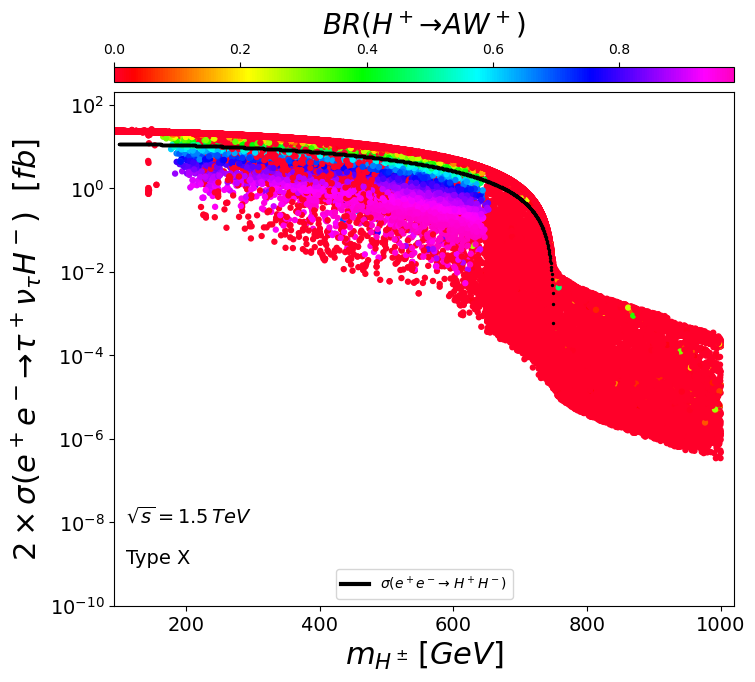}\\
\includegraphics[width=0.3\textwidth]{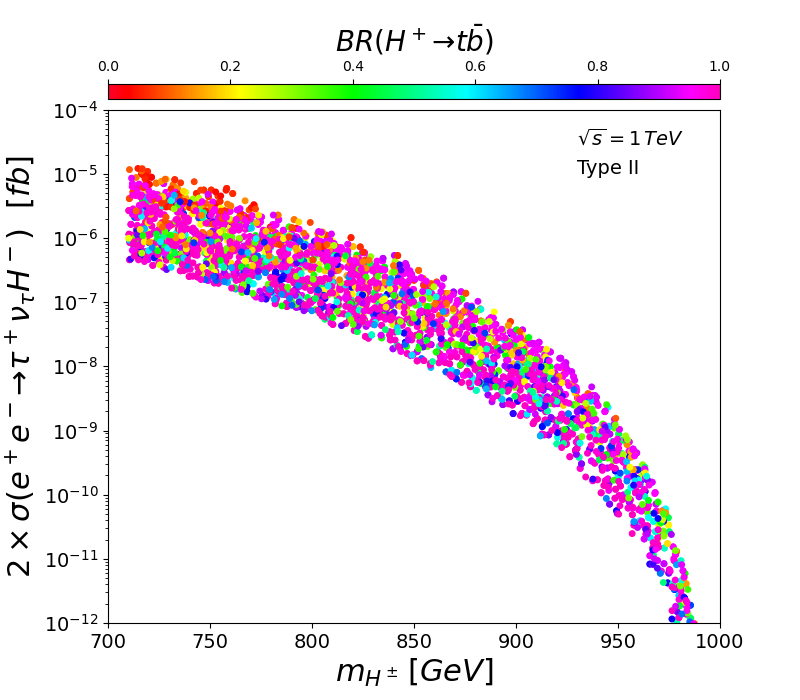}
\includegraphics[width=0.3\textwidth]{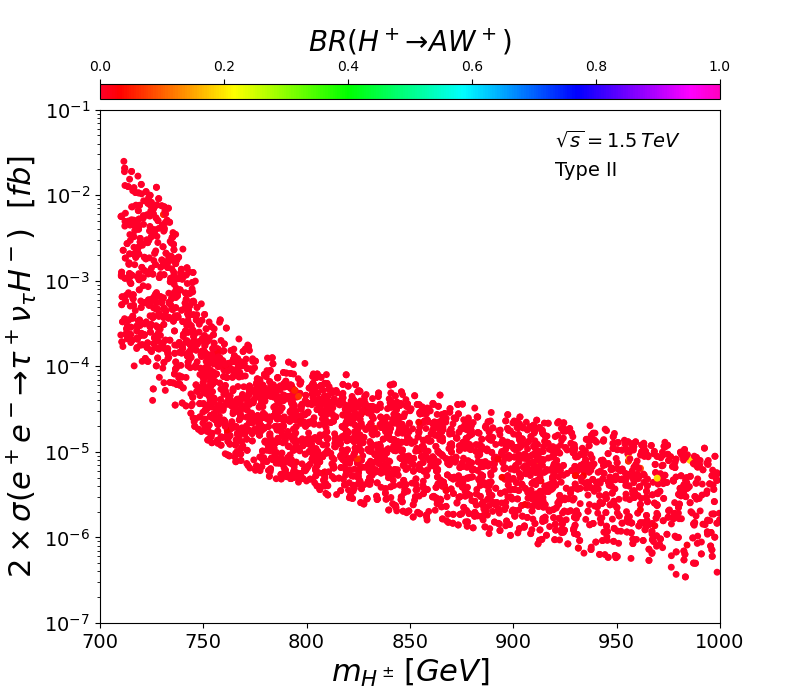}\\
\caption{Total cross section $\sigma(e^+e^- \to \tau^+ \nu_\tau H^-)$ at ILC for $\sqrt{s}=0.5$ TeV, $\sqrt{s}=1$ TeV and  $\sqrt{s}=1.5$ TeV, Type X (upper panel) and Type II (lower panel), as a function of $m_{H^\pm}$. The coding colors indicate one of the following $\br(H^+ \to \tau^+\nu_\tau )$, $\br(H^+ \to t\bar{b} )$ and $\br(H^+ \to HW^+)$ observables. }
\label{fig:fig4}.
\end{figure}

At the ILC/CLIC, the production of a single charged Higgs boson occurs through the processes $e^+ e^- \to \tau^+\nu_\tau H^- $ and $e^+ e^- \to t \bar{b} H^-$, which arise from s-channel interactions involving the exchange of a photon or a $Z$ boson. The s-channel neutral Higgs bosons ($h, H, A$) contributions would be suppressed by the small electron mass. The contribution from t-channel diagrams with $\nu$ exchange is also suppressed by the electron mass.  In the case of the $e^+e^-\to \tau^+\nu_\tau H^-$ process, the dominant contributing diagrams are ($d_1$), ($d_2$) and ($d_3$) of Fig.\ref{fig:fig1}. Similarly for the $t \bar{b} H^-$ process, the dominant contributing diagrams are ($d_1$), ($d_2$), and ($d_3$) of Fig.\ref{fig:fig1}. The coupling of the charged Higgs pair to a photon and Z boson depends only on the electric charge and the Weinberg angle, therefore the cross section for  $e^+ e^- \to \tau^+\nu_\tau H^- $ and $e^+ e^- \to t \bar{b} H^-$ is only sensitive to $\tan\beta$ and charged Higgs mass.

In Fig-\ref{fig:fig4} we plot the total cross section for  $e^+e^- \to \tau^+\nu_\tau H^-$ at ILC for $\sqrt{s}=0.5$ TeV, $\sqrt{s}=1$ TeV and  $\sqrt{s}=1.5$ TeV, Type X (upper panel) and  Type II (lower panel), as a function of  $m_{H^\pm}$. The coding colors indicate one of the following $\br(H^+ \to \tau^{+} \nu_{\tau})$, $\br(H^+ \to t\bar{b} )$ and $\br(H^+ \to H W^+)$ observables. The solid black line corresponds to the total cross section for $e^+e^- \to H^+H^-$. Clearly, in type X, as the center-of-mass energy increases, the cross section decreases significantly. The cross section is primarily dominated by the s-channel diagrams $e^+e^- \to \gamma, Z \to \tau^+ \nu_\tau H^-$, which exhibits the $\frac{1}{s}$ behavior. As the charged-Higgs mass increases from 80 GeV to 250 GeV, the cross section decreases from 170$\fb$ to 1$\fb$  for $\sqrt{s}=500$ GeV. At $\sqrt{s}=0.5, 1, 1.5$ TeV, pair production of charged Higgs is allowed for $m_{H^\pm}< 250, 500, 750$ GeV, respectively, we have checked that the cross section $\sigma(e^+e^- \to \tau^+ \nu_\tau H^-)$ mainly comes from $ \sigma(e^+e^- \to H^+ H^-)\times Br(H^+\to \tau^+ \nu)$ according to the narrow width approximation. After crossing the charged Higgs pair production threshold, one can see that the cross section for $e^+e^- \to \tau^+ \nu_\tau H^-$ drops significantly. It is clear from the plot that the single production of charged Higgs can exceed the pair production slightly in some cases due to the extra Feynman diagrams that contribute to this process.

\begin{figure}[!h]
\centering
\includegraphics[width=0.3\textwidth]{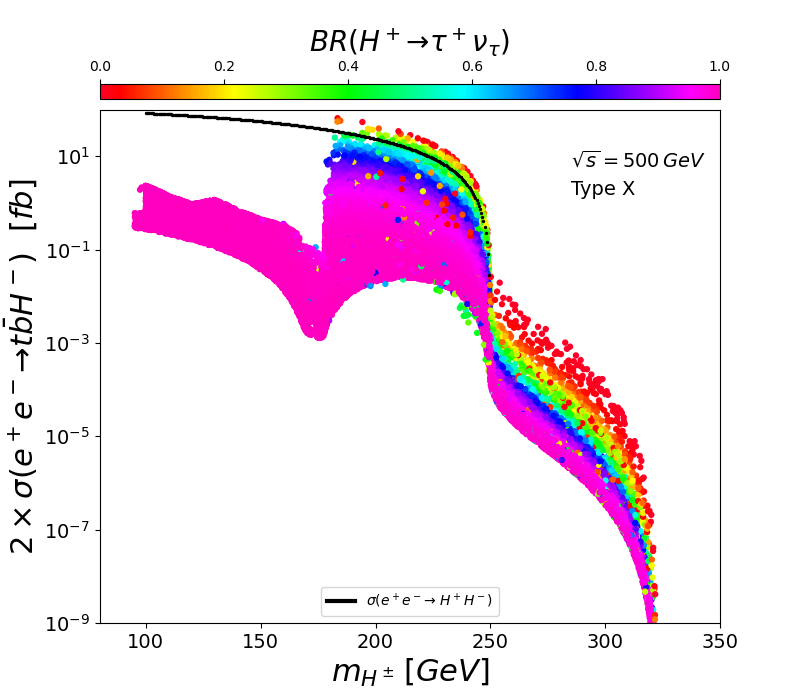}
\includegraphics[width=0.3\textwidth]{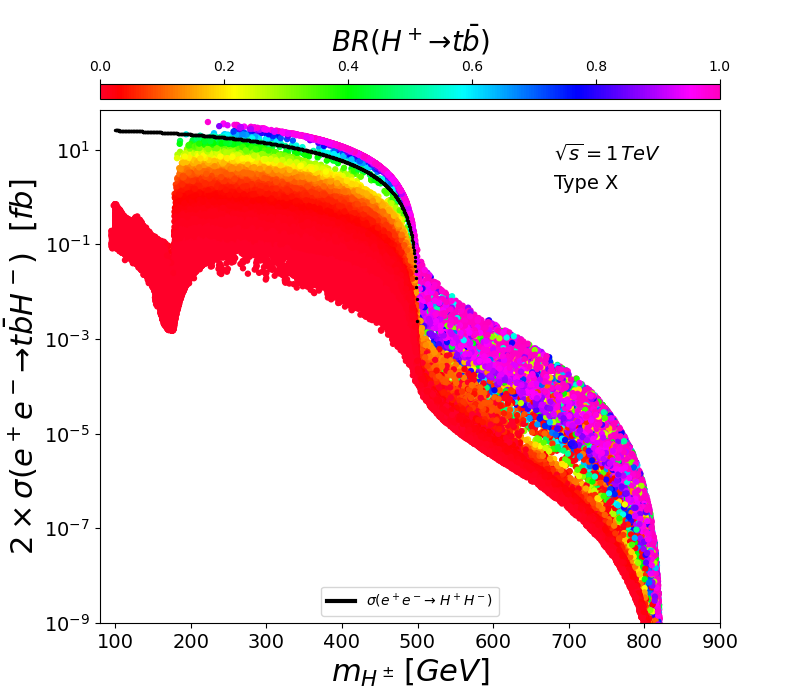}
\includegraphics[width=0.3\textwidth]{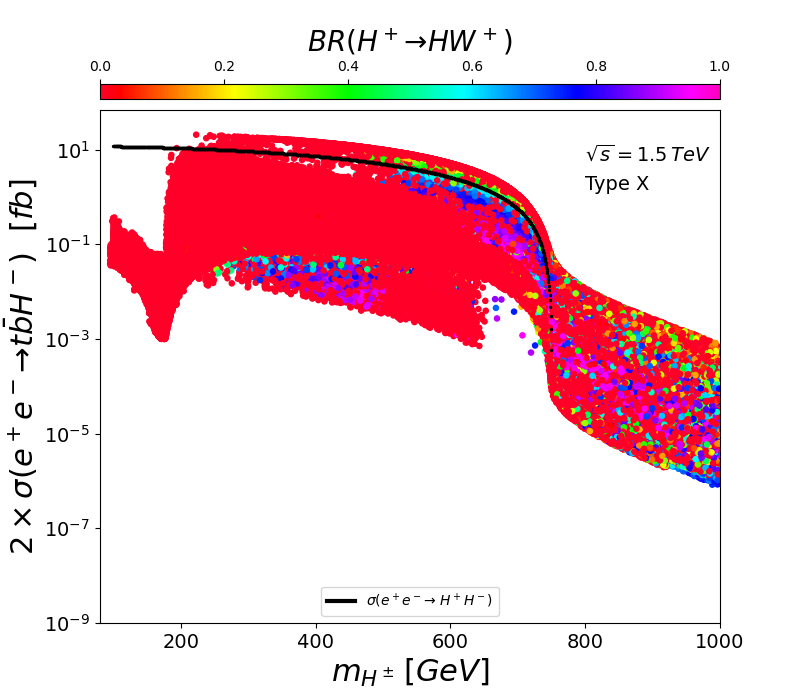}\\
\includegraphics[width=0.3\textwidth]{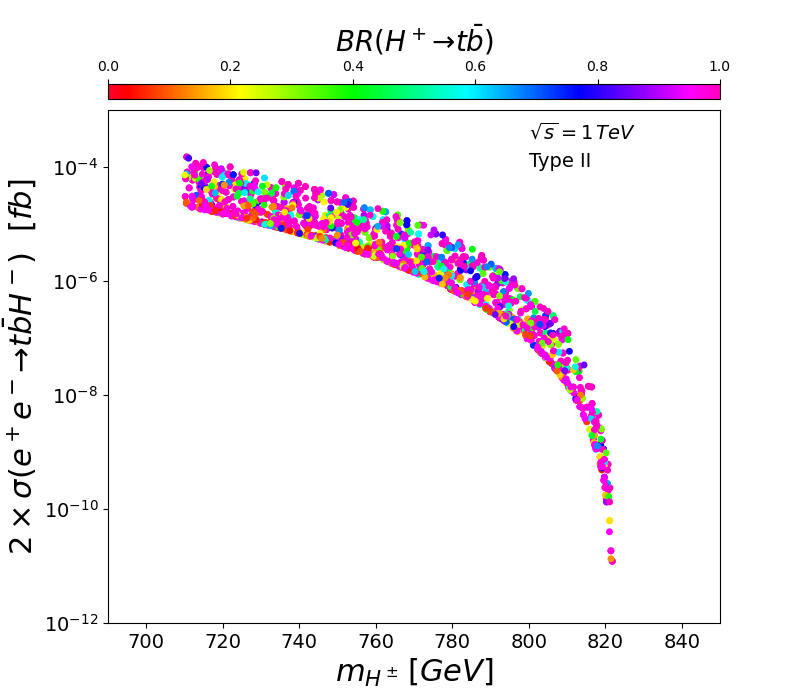}
\includegraphics[width=0.3\textwidth]{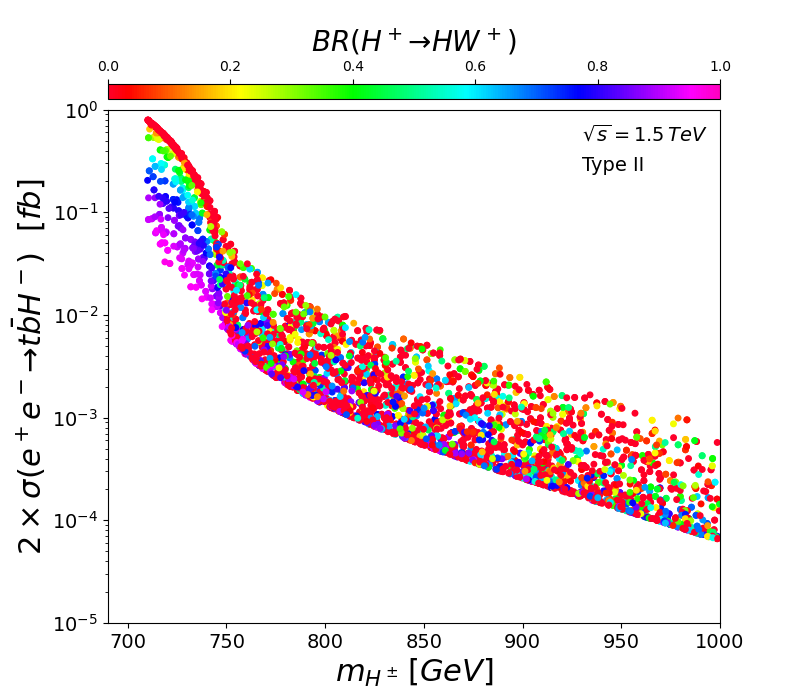}\\
\caption{Total cross section $\sigma(e^+e^- \to t\bar{b}  H^-)$ at ILC for $\sqrt{s}=0.5$ TeV, $\sqrt{s}=1$ TeV and  $\sqrt{s}=1.5$ TeV, Type X (upper panel) and Type II (lower panel), as a function of  $m_{H^\pm}$. The coding colors indicate one of the following $\br(H^+ \to \tau^+\nu_\tau )$, $\br(H^+ \to t\bar{b} )$ and $\br(H^+ \to HW^+)$ observables.}
\label{fig:fig5}
\end{figure}

The cross section for  $e^+e^- \to t\bar{b}  H^-$ at $e^+e^-$ collider with  $\sqrt{s}=0.5$ TeV, $\sqrt{s}=1$ TeV and $\sqrt{s}=1.5$ TeV  are illustrated in Fig-\ref{fig:fig5} as a function of  $m_{H^\pm}$. The coding colors indicate one of the following  $\br(H^+ \to W^+H)$,  $\br(H^+ \to \tau^+\nu_\tau )$ and $\br(H^+ \to t\bar{b} )$  observables. The solid black line corresponds to the total cross section for $e^+e^- \to H^+H^-$. For type X, the cross section increases significantly as $m_{H^+}$ approaches $m_t + m_b$, primarily due to the production  via $\sigma(e^+ e^- \to H^+ H^-)\times \mathrm{BR}(H^+ \to t\bar{b})$. It then decreases for large $m_{H^+}$. In type X, the charged Higgs coupling to $tb$  is proportional to $1/\tan\beta$ the cross section for $e^+e^- \to t\bar{b}  H^-$ is then suppressed for large $\tan\beta$. In this case as well, one can see that $\sigma(e^+e^- \to t\bar{b}  H^-)$ is a bit higher than the pair production of charged Higgs   $\sigma(e^+e^- \to H^+ H^-)$.

For Type II, as the center-of-mass energy increases, the cross section increases substantially. The allowed parameter space for $\tan \beta <$ 10 and a charged Higgs mass of $m_{H^+} >$ 700GeV shows that the cross section depends on the charged Higgs mass. It reaches a maximum value of approximately 0.9$\fb$ at $\sqrt{s}=1.5$ TeV before starting to decrease as $m_{H^+}$  increases. 

\begin{figure}[!h]
\centering
\includegraphics[width=0.3\textwidth]{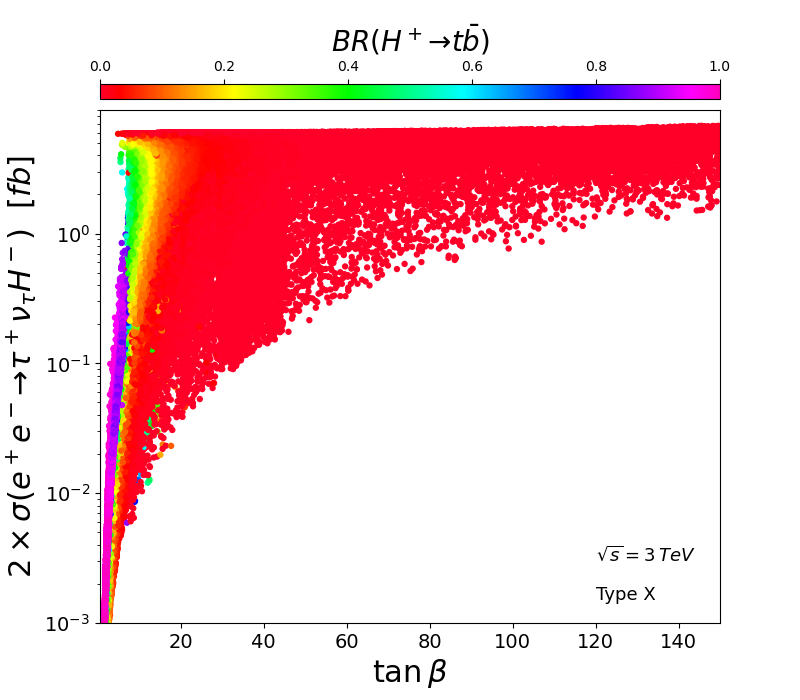}
\includegraphics[width=0.3\textwidth]{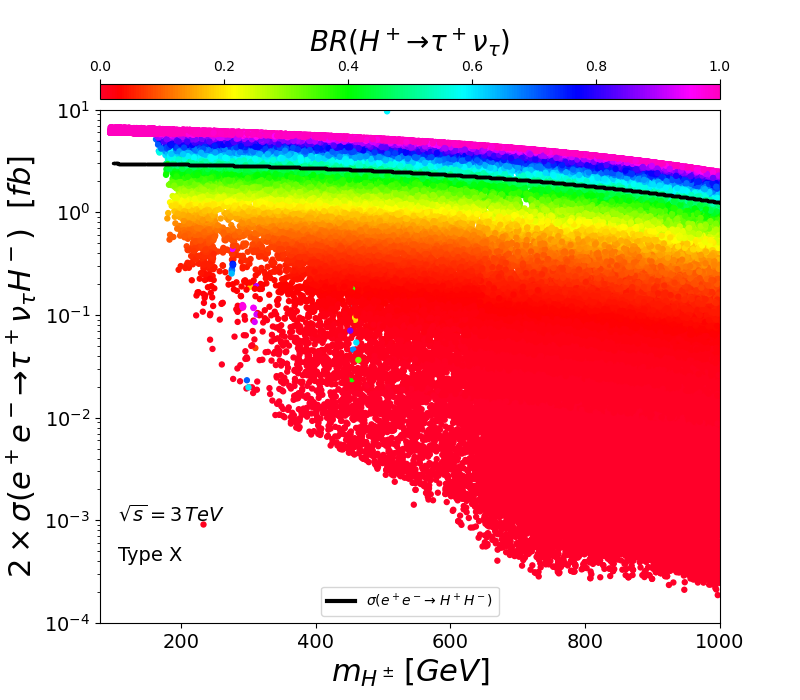}
\includegraphics[width=0.3\textwidth]{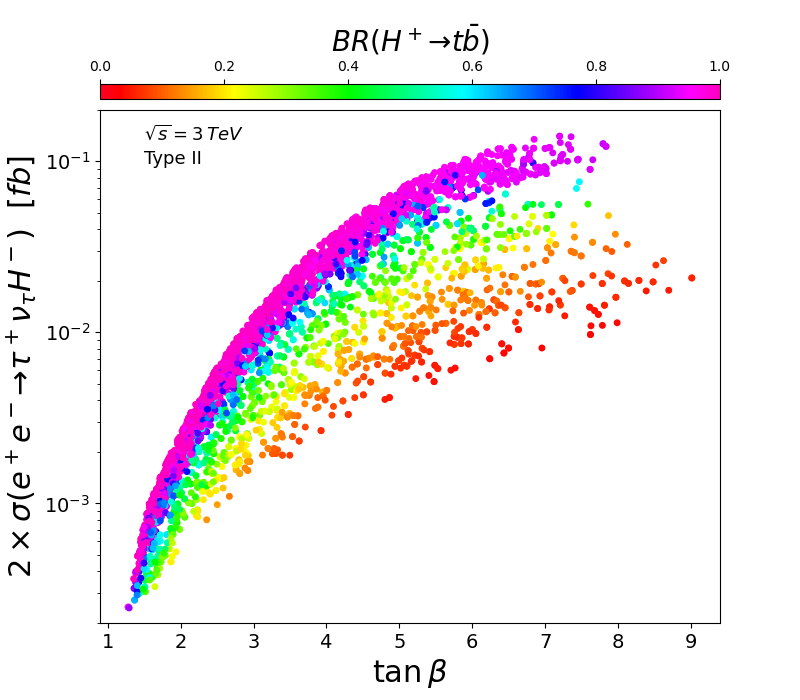}\\
\includegraphics[width=0.3\textwidth]{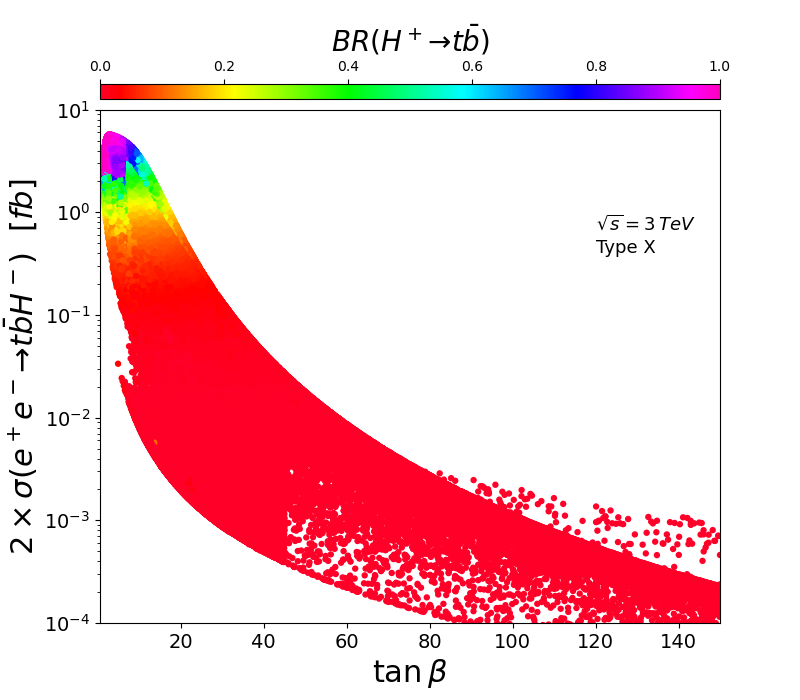}
\includegraphics[width=0.3\textwidth]{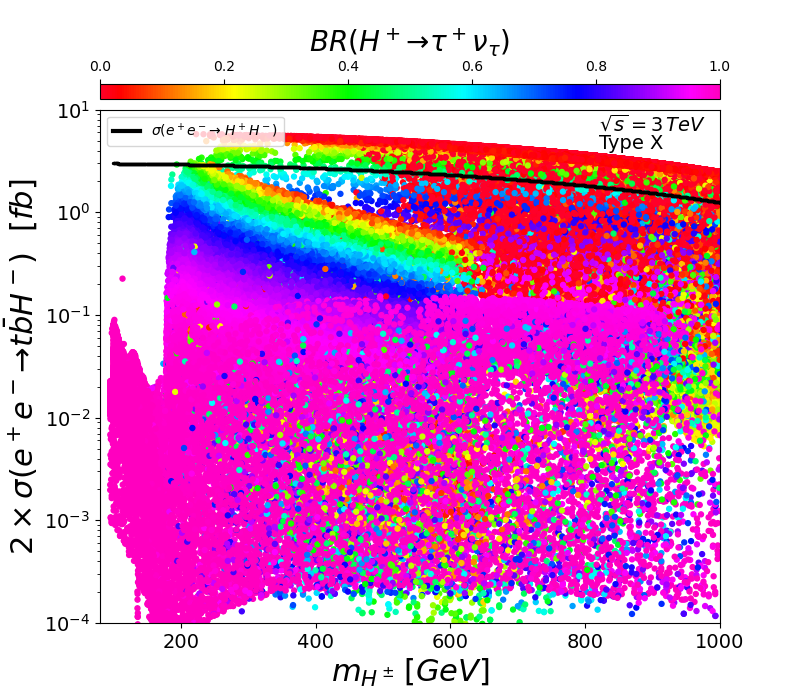}
\includegraphics[width=0.3\textwidth]{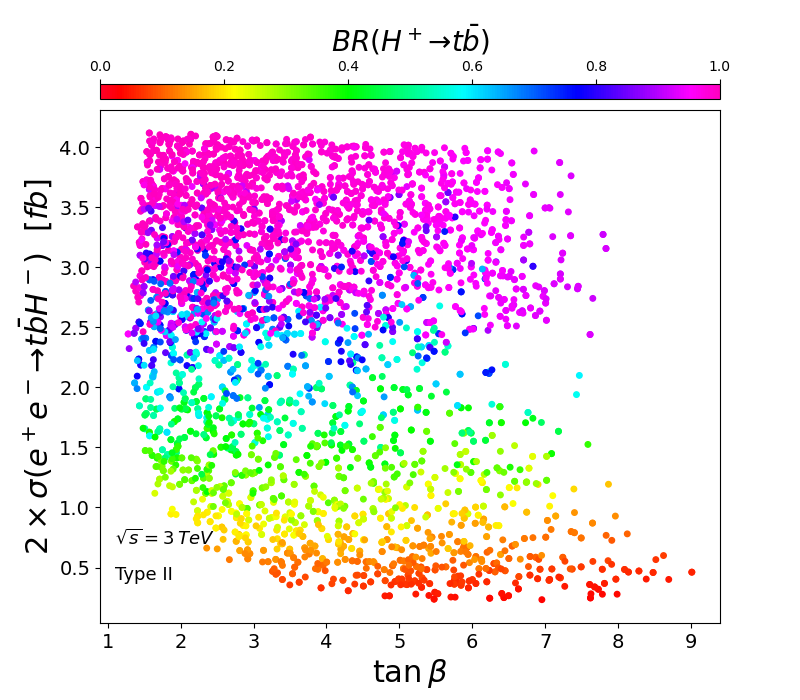}\\
\caption{Total cross section $\sigma(e^+e^- \to \tau^+ \nu_\tau H^-)$ (upper panel) and $\sigma(e^+e^- \to t\bar{b}  H^-)$ (lower panel) at CLIC for  $\sqrt{s}=3$ TeV, Type II and Type X, as a function of  $m_{H^\pm}$ and $\tan\beta$. The coding colors indicate one of the following   $\br(H^+ \to \tau^+\nu_\tau )$ and $\br(H^+ \to t\bar{b} )$ observables. }
\label{fig:fig6}
\end{figure}

At $\sqrt{s}$= 3 TeV the cross section shows a slight overall improvement in the permitted dataset as drawn in Figure \ref{fig:fig6}. It can reach a maximum value - up to 7$\fb$ for lower $m_{H^+}$ and large tan$\beta$.  As the charged-Higgs mass increases from 80 GeV to 1 TeV, the cross section  decreases from 7$\fb$ to 1.5$\fb$. However, for type II the cross section can reach a peak value of up to 0.14$\fb$  for $m_{H^+}$ larger than 700 GeV and lower $\tan \beta $. We also know that, the large $\tan \beta $ parameter space region is not allowed by LHC Higgs data (as seen in lower right panel). Keep in mind that the 2HDM type II and type-X are both affected by the enhancement for large
tan $\beta$  (see Table \ref{coupIII} for the couplings).

In the Figure-\ref{fig:fig4}, one can also read branching fractions of the charged Higgs.  For Type X,  the decay mode  $\br(H^+ \to \tau^+\nu_\tau )$ dominates. However, as observed in the lower left panel, the decay mode $\br(H^+ \to t\bar{b} )$ dominates for type II, a
significant competition arises among the decay modes $\br(H^+ \to W^+A)$, $\br(H^+ \to W^+h)$ and $\br(H^+ \to W^+H)$.

\subsection{$e^+ e^- \to W^\pm H^\mp Z$ and $e^+ e^- \to W^\pm H^\mp S$, $S=h,H,A$,  processes}
\begin{figure}[!ht]
\centering
\includegraphics[width=0.31\textwidth]{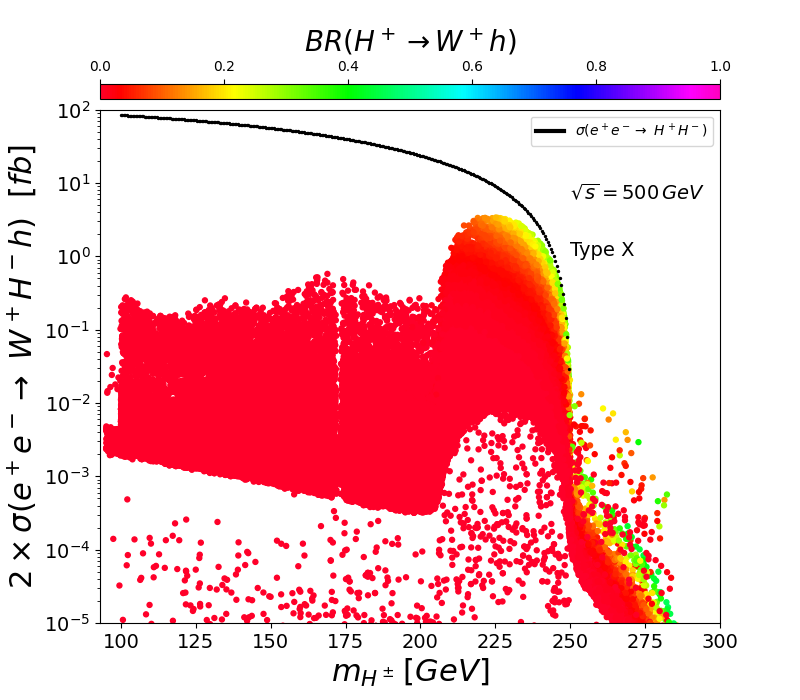}
\includegraphics[width=0.31\textwidth]{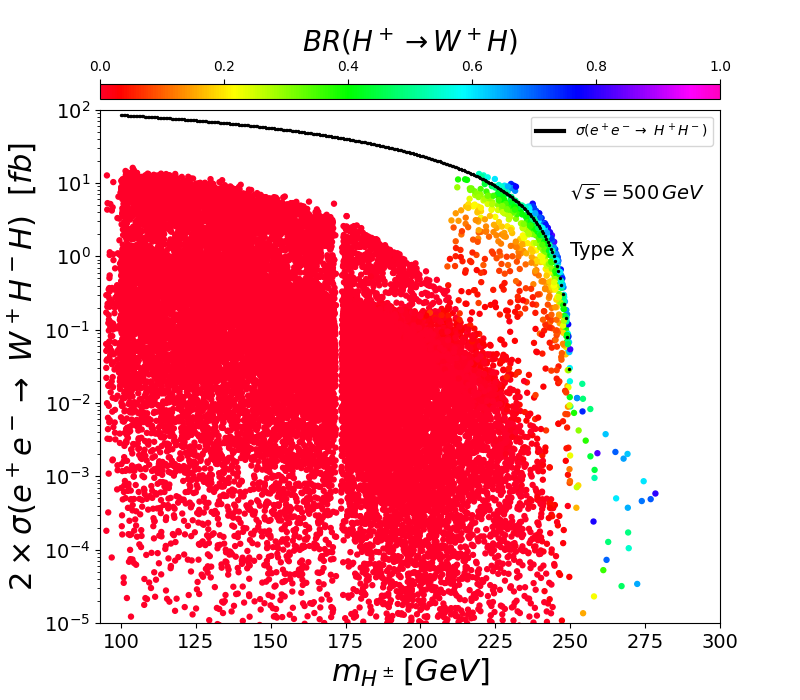}
\includegraphics[width=0.31\textwidth]{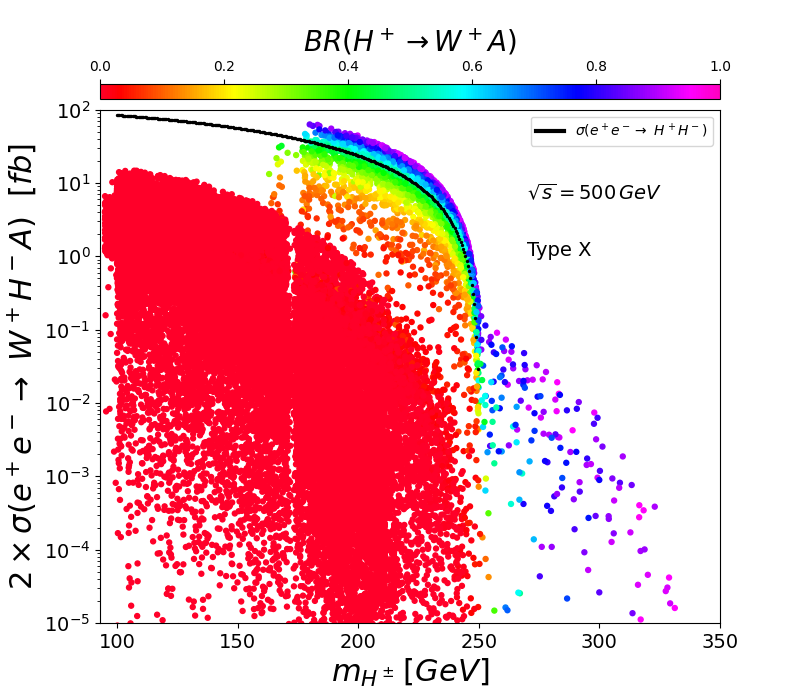}\\
\includegraphics[width=0.31\textwidth]{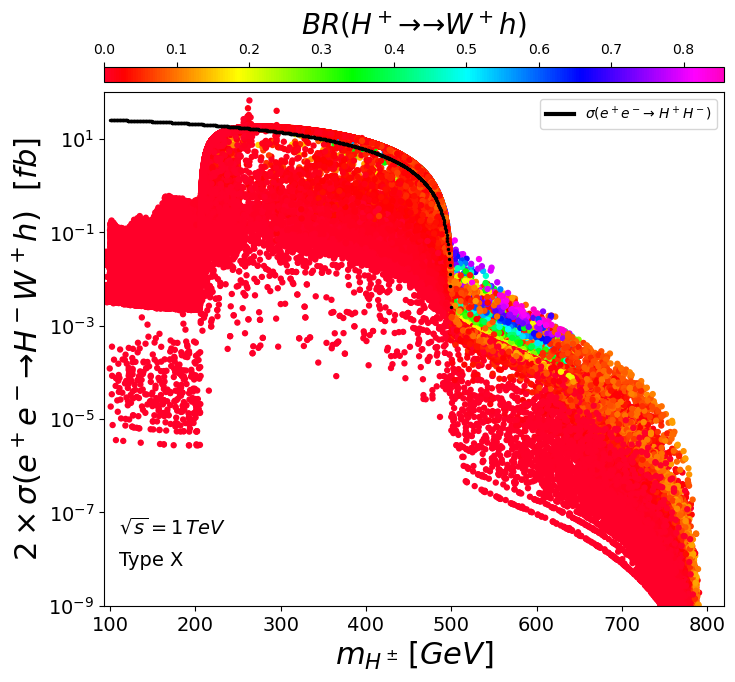}
\includegraphics[width=0.31\textwidth]{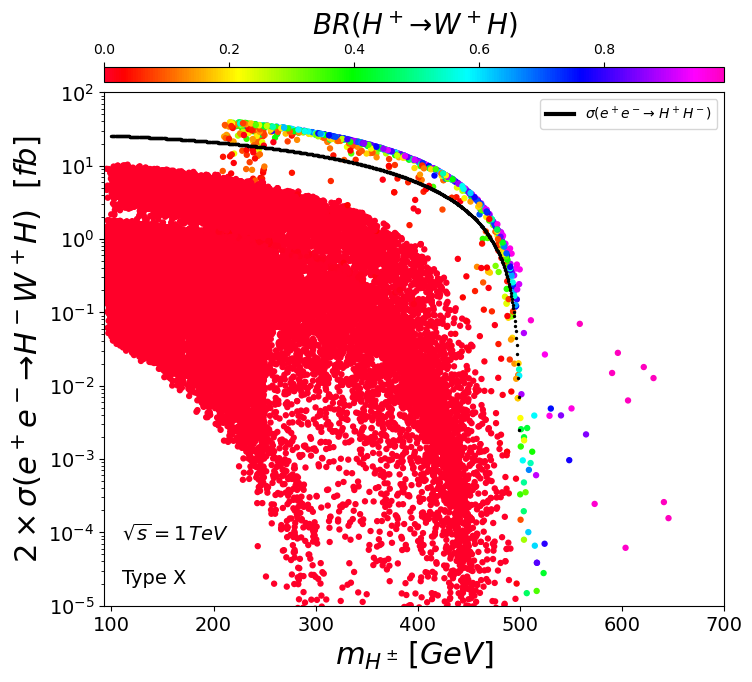}
\includegraphics[width=0.3\textwidth]{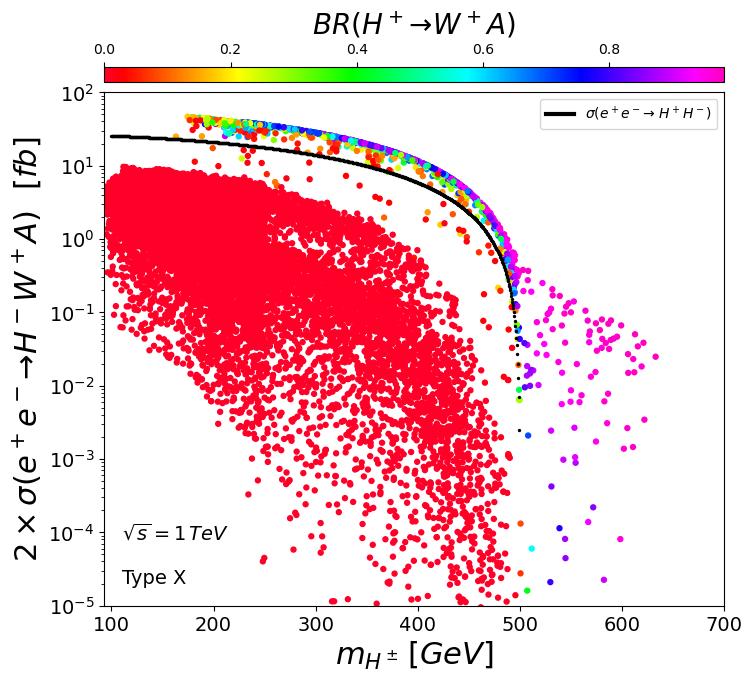}\\
\includegraphics[width=0.31\textwidth]{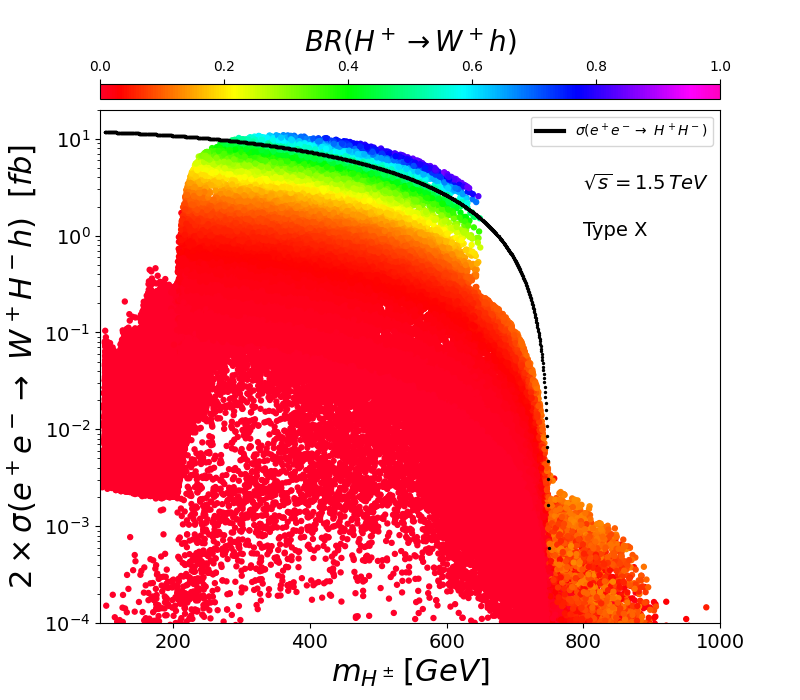}
\includegraphics[width=0.31\textwidth]{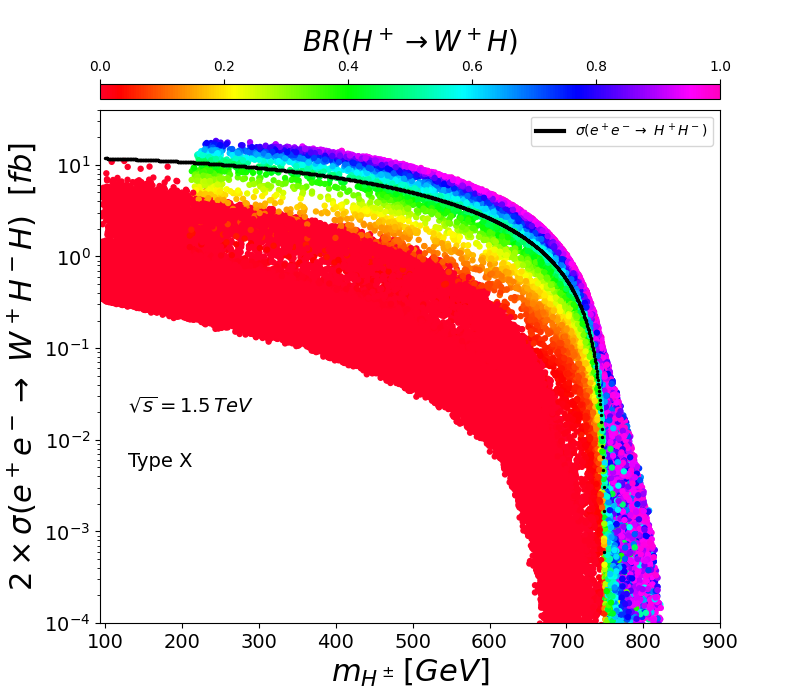}
\includegraphics[width=0.31\textwidth]{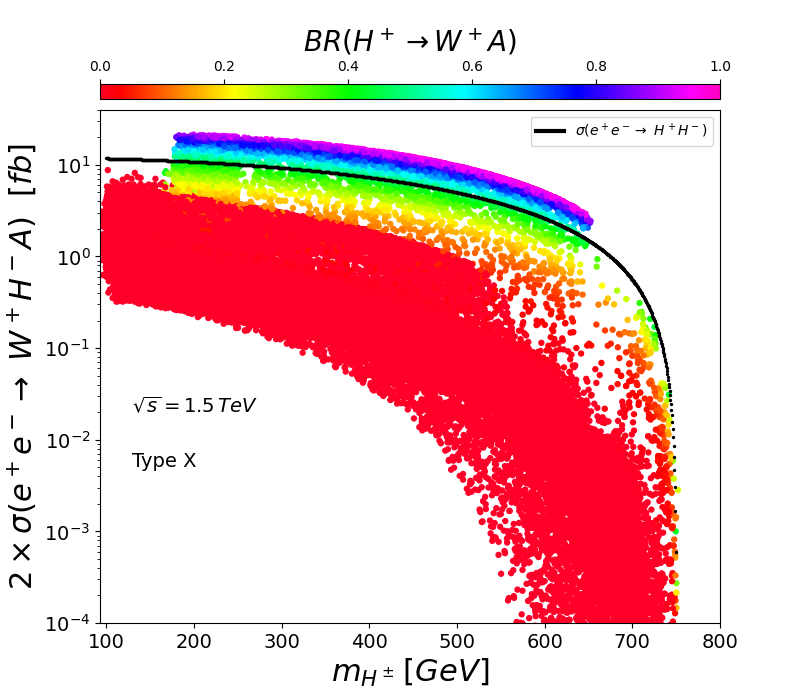}\\
\caption{Total cross section $\sigma(e^+e^- \to { W^\pm H^\mp h\, , \ W^\pm H^\mp H\ ,\, W^\pm H^\mp A} )$ at ILC for $\sqrt{s}=500$ GeV (upper panel), $\sqrt{s}=1$ TeV (middle panel) and  $\sqrt{s}=1.5$ TeV (lower panel) for 2HDM all types, as a function of either $m_{H^\pm}$. The coding colors indicate one of the following $\br(H^+ \to hW^+)$,  $\br(H^+ \to HW^+)$, $\br(H^+ \to AW^+)$ observables. }
\label{fig:fig7}.
\end{figure}
Prior to discussing our findings for $e^+e^- \to H^\pm W^\mp Z$  and $e^+e^- \to H^\pm W^\mp S$, it is important to note that these processes, at $e^+e^-$ collider, are not sensitive to the 2HDM's Yukawa structures. Therefore, the cross sections we will present are the same for all Yukawa types of the 2HDM. However, the computation of the branching fractions is model dependent. Our results for branching fractions are shown for 2HDM type X.

As previously stated, the cross section for $e^+e^- \to W^\pm H^\mp Z$ is proportional to $s^2_{\beta-\alpha} c^2_{\beta-\alpha}$. Since  $c_{\beta-\alpha}$ is constrained by LHC data to be rather small ( $0.003< c^2_{\beta-\alpha}<0.063$ ), it is expected that $\sigma(e^+e^- \to W^\pm H^\mp Z)$ would be suppressed. Numerically, we have checked that this cross section could at most reach $10^{-2}$$\fb$ for the center of mass energy 500 GeV, 1 TeV, and 3 TeV. We will not show any numerical outcomes for this process. In Fig.\ref{fig:fig7}, we illustrate cross sections for $e^+e^- \to {W^\pm H^\mp h\, , \ W^\pm H^\mp H\ ,\, W^\pm H^\mp A}$ for $\sqrt{s}=0.5$ TeV (upper panels), 1 TeV (middle panels) and 1.5 TeV (lower panels). After a brief review of the Feynman rules, we find that all the couplings that involve $h$ (resp H) and contribute to $e^+e^- \to W^\pm H^\mp h$  (resp $e^+e^- \to W^\pm H^\mp H$ ) are proportional to $c_{\beta-\alpha}$ (resp. $s_{\beta-\alpha}$). The majority of Feynman diagrams in the $e^+e^- \to W^\pm H^\mp A$ scenario lack both the $s_{\beta-\alpha}$ and $c_{\beta-\alpha}$ mixing factors, this is because $AW^\pm H^\mp$ is proportional to $g/2$.
Only Feynman diagram  $d_5$ would have $c_{\beta-\alpha}$ (resp $s_{\beta-\alpha}$)  factor  coming from $e^+e^-\to Ah^* \to AW^\pm H^\mp$ (resp $e^+e^-\to AH^* \to AW^\pm H^\mp$). The cross section for $e^+e^- \to W^\pm H^\mp h$ is, as predicted, somewhat smaller than the cross sections for $e^+e^- \to W^\pm H^\mp H$ and $e^+e^- \to W^\pm H^\mp A$. The reason is that $e^+e^- \to W^\pm H^\mp h$ is proportional to $c^2_{\beta-\alpha}$ which is rather small. At $\sqrt{s}=500$ GeV, for $m_{H^\pm}<205$ GeV, where $H^+\to Wh$ is closed, $\sigma(e^+e^- \to W^\pm H^\mp h)$ is at the level of 0.1$\fb$. The cross section exhibits some increase once we cross the $H^+ \to W^+h$ threshold and $Br(H^+ \to W^+ h)$ becomes significant (see the horizontal rectangle with color coding). The bump on the cross section of $e^+e^-\to  W^\pm H^\mp h$ is due to the on-shell production of a pair of charged Higgs followed by  $H^+ \to W^+h$ decay: $\sigma(e^+e^-\to  W^\pm H^\mp h) \approx \sigma(e^+e^-\to H^+ H^-) \times Br( H^+\to  W^+h)$.
When charged Higgs pair production is kinematically prohibited for $m_{H^\pm}>250$ GeV, the cross section dramatically decreases to $10^{-3}$$\fb$.\\
In the case of  $e^+e^- \to W^\pm H^\mp H$ production, which is proportional to $s^2_{\beta-\alpha}$, one can see that the size of this cross section is slightly higher than for  $e^+e^- \to W^\pm H^\mp h$. The cross section may reach 10$\fb$ before the opening of $H^+\to W^+ H$, and it would drop as the charged Higgs mass increases. Once we cross the $H^+ \to W^+H$ threshold, one can see an enhancement of the cross section coming from the resonant production originating from $H^+\to W^+ H$: $\sigma(e^+e^- \to W^\pm H^\mp H)\approx \sigma(e^+e^- \to H^+H^-) \times Br(H^+ \to W^+ H)$.\\
The cross sections for $e^+e^- \to W^\pm H^\mp A$, size and behavior, are rather similar to those of $e^+e^- \to W^\pm H^\mp H$. The resonant production in this case would come from the $H^+\to W^+ A$ decay:  $\sigma(e^+e^- \to W^\pm H^\mp A)\approx  \sigma(e^+e^- \to H^+H^-) \times Br(H^+\to W^+ A)$.\\
At $\sqrt{s}=1$ TeV or (resp 1.5 TeV), we observe a similar pattern as for 500 GeV. In this case, we have more phase space for charged Higgs pair production: $m_{H^\pm}<500$ GeV (resp  $m_{H^\pm}<750$ GeV) and this is the reason why one can see that the bump of the resonant production is slightly larger than in the previous case.

\begin{figure}[!h]
\centering
\includegraphics[width=0.31\textwidth]{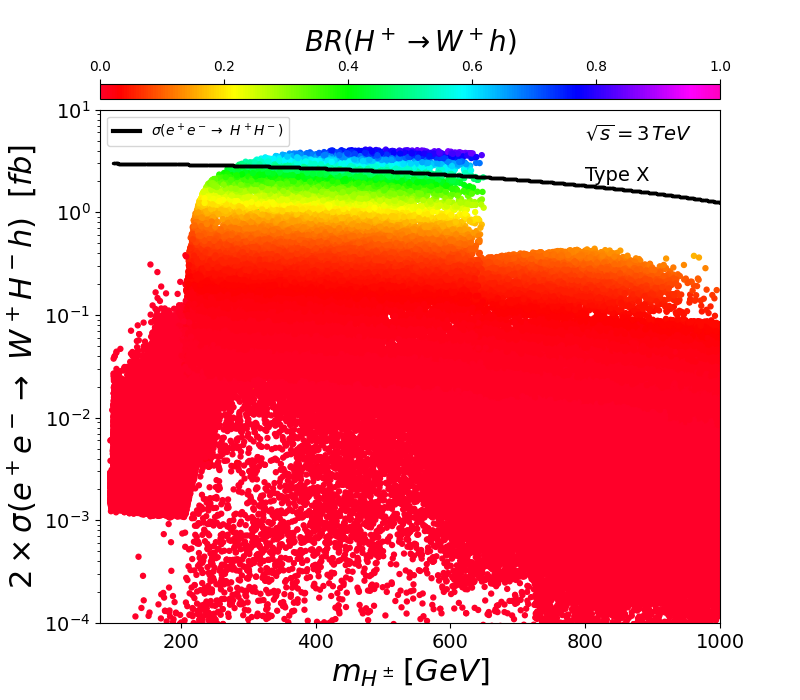}
\includegraphics[width=0.31\textwidth]{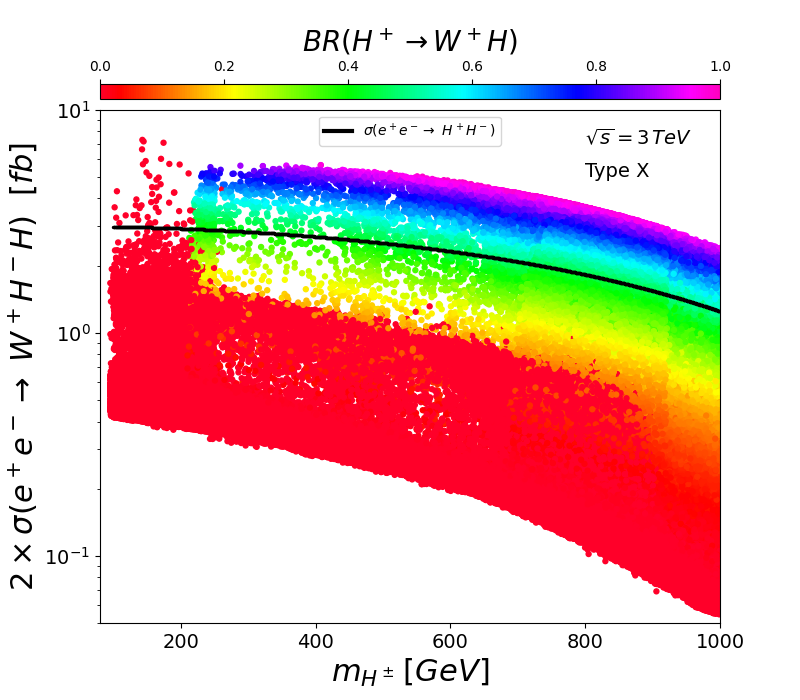}
\includegraphics[width=0.31\textwidth]{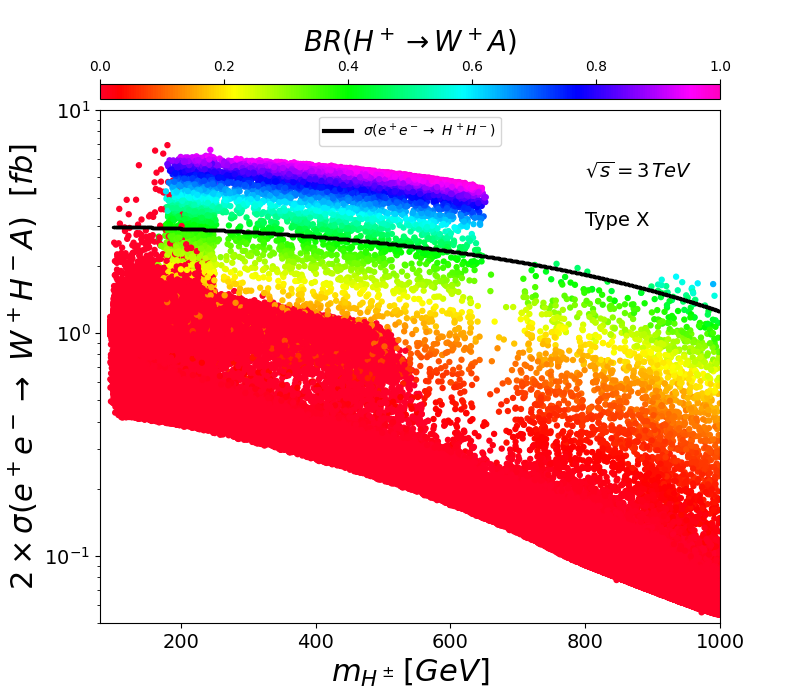}\\
\caption{Total cross section $\sigma(e^+e^- \to {W^\pm H^\mp h\, , \ W^\pm H^\mp H\ ,\, W^\pm H^\mp A} )$ at CLIC for  $\sqrt{s}=3$ TeV for 
2HDM all types, as a function of either $m_{H^\pm}$. 
The coding colors indicate one of the following $\br(H^+ \to hW^+)$,  $\br(H^+ \to HW^+)$, $\br(H^+ \to AW^+)$ observables. }
\label{fig:fig8}.
\end{figure}
Now we move to the high energy case: $\sqrt{s}=3$ TeV. The cross sections are illustrated in Fig.\ref{fig:fig8}. As one can see from the left panel, away from the resonant production, the cross section for $e^+e^- \to W^\pm H^\mp h$ is rather small: less than 0.3$\fb$. In the region where $Br(H^+ \to W^+ h)$ is sizable one can see an enhancement of the cross section that can reach the level of 1$\fb$.  As previously mentioned, the cross section is tiny because of its its proportionality to $c^2_{\beta-\alpha}$. In the case of $e^+e^- \to W^\pm H^\mp H$ and $e^+e^- \to W^\pm H^\mp A$, we do not have such a suppression factor and also the t-channel contribution tends to enhance the total cross section for such higher center of mass energy of 3 TeV. Therefore, it is evident that the total cross section is larger than 0.1$\fb$ throughout the whole range of charged Higgs mass. 
In this context, it is worth noting that although t-channel contributions are equally small for all processes, they remain the subject of intense interference with the s-channel contributions. This interference, which can significantly affect the overall cross section, is highly sensitive to the specific couplings involved in each process, and depends also on whether the thresholds $H^+ \to W^+h(H)$ are kinematically open or not.
In all panels, we show on the horizontal rectangle with color coding the branching ratios $Br(H^+\to W^+ h)$ or $Br(H^+\to W^+ H)$ or $Br(H^+\to W^+ A)$. 

\subsection{$\mu^+ \mu^- \to \tau^+ \nu_\tau H^- $ }
We now present our results at the muon collider with 3 TeV center of mass energy. The main difference with respect to the case of $e^+e^-$ collider is the sensitivity to the diagrams with s-channel Higgs exchange like $(d_{6,7,8})$ and t and u channel of Fig.\ref{fig:fig1}. As discussed before, in 2HDM type X, both neutral and charged Higgs couplings to a pair of fermions are proportional to $\tan\beta$, the amplitude of the Feynman diagram like $(d_8)$-Fig.\ref{fig:fig1} behaves like $\tan^3\beta$ which could lead to a factor of $\tan^6\beta$ in the square of amplitude. This $\tan^6\beta$ factor could also originate from diagram $(d_6)$-Fig.\ref{fig:fig1}, since both $hH^+H^-$ and $HH^+H^-$ couplings contains respectively : $(m_h^2 - 2\frac{m_{12}^2}{s_{2\beta}} \Big)\frac{c_{\beta+\alpha}}{s_\beta c_\beta} $ and $(m_H^2 - 2\frac{m_{12}^2}{s_{2\beta}} \Big)\frac{c_{\beta+\alpha}}{s_\beta c_\beta} $ which both scales like $\tan\beta$ at large $\tan\beta$ limit.

\begin{figure}[!h]
\centering
\includegraphics[width=0.4\textwidth]{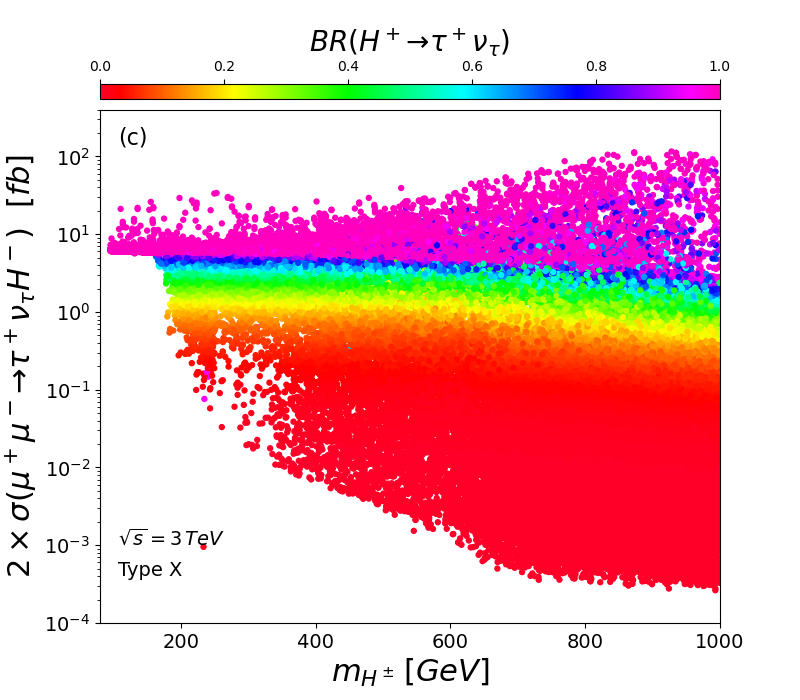}
\includegraphics[width=0.4\textwidth]{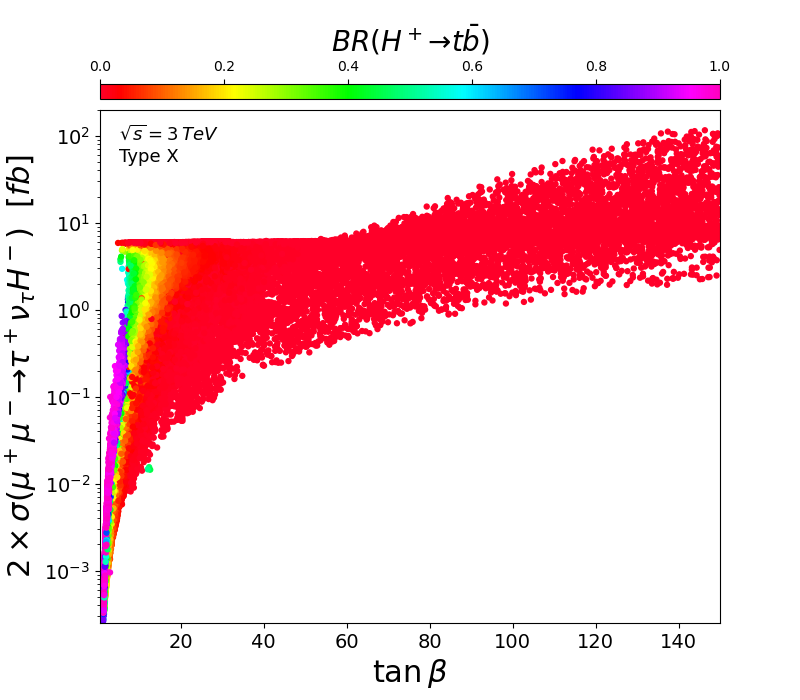}\\
\caption{Total cross section $\sigma(\mu^+\mu^- \to \tau^+ \nu_\tau H^-)$ at MuC for $\sqrt{s}=3$ TeV in 2HDM Type X, as a function of either $m_{H^\pm}$ and $\tan\beta$. The coding colors indicate one of the following $\br(H^+ \to \tau^+\nu_\tau )$ and $\br(H^+ \to t\bar{b} )$ observables. 
}
\label{fig:fig9}
\end{figure}

In Fig.\ref{fig:fig9}, we illustrate the total cross section $\sigma(\mu^+\mu^- \to \tau^+\bar{\nu}_\tau H^-)$ at a muon collider for  center-of-mass energy $\sqrt{s}=$ 3 TeV. The cross sections are examined as a functions of $\tan\beta$ and the charged Higgs boson mass ($m_{H^\pm}$). The plots are color-coded to indicate specific branching ratios for different Higgs decay channels: $Br(H^+ \to \tau^+\nu_\tau)$ and $Br(H^+ \to t\bar{b})$. At first glance, one can notice that the allowed parameter space significantly enhances the total cross section of this process in the large $\tan\beta$ limit. In the right column of Fig.\ref{fig:fig9}, the cross sections are plotted as a function of $\tan\beta$ in Type X.  For 3 TeV center of mass energy, the cross section shows an increase for  $\tan\beta$ in the range 60 to 150, reaching 117 \fb. This enhancement at large $\tan\beta$ values is due to the factor $\tan^6\beta$ that we have explained above.

In 2HDM type X, as can be seen from the left column of Fig.\ref{fig:fig9},  
the cross section reaches its maximum for charged Higgs mass above 700 GeV. The significant increase in the cross section with increasing $\tan\beta$ from 1 to 9, reaching 0.14$\fb$, can be explained by the behavior of the charged Higgs boson coupling to leptons $g_{H^+\tau^-\bar{\nu}_{\tau}}$,  which is proportional to $ \tan\beta$  (similarly for the 2HDM Type X). The cross sections for the 2HDM Type X are generally larger than those of the 2HDM Type II. We emphasize that the theoretical and experimental constraints still allow for $\tan\beta \geq 150$ in type X, whereas in 2HDM type II the allowed value is $\tan\beta \leq 10$ unless if we are very close to the decoupling limit $c_{\beta-\alpha}\approx 0$. 
In our study, we only scan up to $\tan\beta \leq 150$. This suggests that the cross sections in  type X are generally larger than those in type II.

Furthermore, it can be seen from Fig.\ref{fig:fig6} and  Fig.\ref{fig:fig9} that, in type X, 
the cross section $\mu^+ \mu^- \to \tau^+ \nu_\tau H^-$ is ten times greater than the corresponding cross section 
$e^+ e^- \to \tau^+ \nu_\tau H^-$ at $e^+e^-$ colliders for the same center-of-mass $\sqrt{s}$ and for large $\tan\beta$. However, for $\tan\beta \le$ 60,  the cross section is the same both for $\mu^+\mu^-$ and $e^+e^-$ colliders $\sigma(\mu^+ \mu^- \to \tau^+ \nu_\tau H^-)$=$\sigma(e^+ e^- \to \tau^+ \nu_\tau H^-)$. In contrast, in Type II, the cross sections are the same in both colliders.

The CP-even Higgs can decay into: $b\bar{b}$, $\tau^+ \tau^-$, $WW$, $ZZ$, $t\bar{t}$, $ZA$, $hh$, $W^\mp H^\pm$, and $H^+ H^-$. In the 2HDM type X, $H \to \tau^+ \tau^-$ would be the dominant decay mode at large $\tan \beta$. $H \to WW$, $ZZ$ are suppressed since both are proportional to $\cos(\beta - \alpha) \approx 0$, but could nevertheless reach a few percent branching fraction in some cases. After crossing the $t\bar{t}$ threshold, there is a strong competition among $hh$, $ZA$, $W^\pm H^\mp$, and $H^+ H^-$. The decay channel $H \to H^+ H^-$, which is open only for $m_H > 2m_{H^\pm}$, is rather small compared to $H \to W^\pm H^\mp$ and $H \to ZA$, which have larger phase space and the coupling $H W^\pm H^\mp \propto \sin(\beta - \alpha)$ is maximal and this makes the $\text{Br}(H \to W^\mp H^\pm)$ rather substantial as can be seen from Fig.\ref{fig:fig9}.

\subsection{$\mu^+\mu^- \to t \bar{b} H^-$ }
We will now discuss the process $\mu^+ \mu^- \to t \bar{b} H^-$. This process is primarily governed by two prominent Feynman diagrams: Fig.\ref{fig:fig1}-($d_6$) that involves the particles $S = h,\,H,\,A$, and Fig.\ref{fig:fig1}-($d_9$), where $S_k = H^\pm$. In the context of the 2HDM type-X, such Feynman diagrams involve three vertices of what is directly proportional to the $\tan\beta$.  While, in the coupling $t \bar{b} H^-$, we have $\tan\beta$ enhancement for the bottom mass and $1/\tan\beta$ suppression for the top mass. This makes the $\tan\beta$ behavior different for $\mu^+ \mu^- \to t \bar{b} H^-$ compared to $\mu^+ \mu^- \to \tau^+ \nu_\tau H^-$.

\begin{figure}[!ht]
\centering
\includegraphics[width=0.4\textwidth]{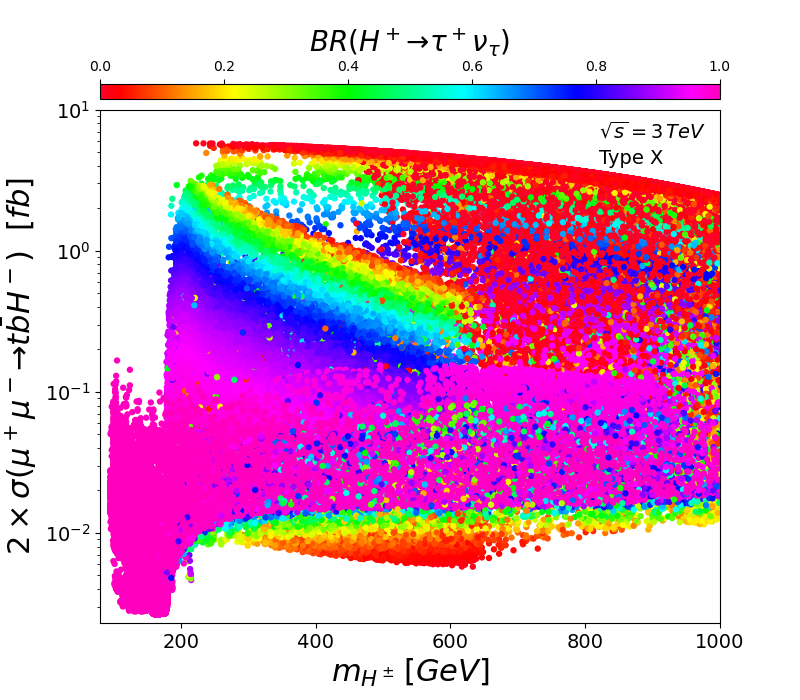}
\includegraphics[width=0.4\textwidth]{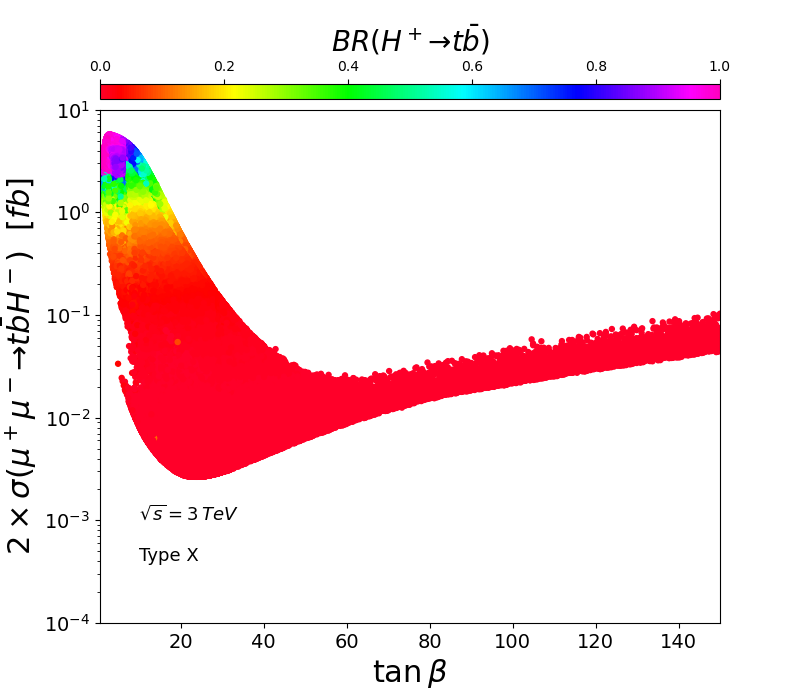}\\
\caption{Total cross section $\sigma(\mu^+\mu^- \to t \bar{b} H^-)$  at muon collider for $\sqrt{s}=3$ TeV in 2HDM Type X, 
as a function of either $m_{H^\pm}$ and $\tan\beta$. The coding colors indicate one of the following $\br(H \to \tau^+\nu_\tau )$ and $\br(H^+ \to t\bar{b})$ 
observables.}
\label{fig:fig10}
\end{figure}

In Fig.\ref{fig:fig10}, we illustrate the total cross section $\sigma(\mu^+\mu^- \to t \bar{b} H^-)$ as a functions of the charged Higgs mass $m_{H^+}$ and $\tan \beta$ for 2HDM type X and II at $\sqrt{s}=3$ TeV. The charged Higgs boson mass, $m_{H^+}$, plays a crucial role in the variation of the cross section. In type X, for $m_{H^+} \lesssim 170$ GeV, the cross section is distinctly lower. This production primarily arises from the top-pair production process followed by one top decay into $H^+$ and a bottom quark: $\sigma(\mu^{+} \mu^{-} \rightarrow t\bar{t}) \times Br(t \rightarrow H^{-} b )$. For  $m_{H^+} \sim 171$ GeV the cross section is relatively small and starts to rise, reaching maximum values of approximately $5.8 \ \fb$. This increase is primarily due to the opening of channels involving:   either $H \rightarrow W^{\pm *}H^{\mp}$ followed by $W^{+ *} \rightarrow t \bar{b}$  or  $\mu^+\mu^- \to \{ \gamma, Z, H,h\}  \rightarrow H^{+}H^{-}$ followed by $H^{+} \rightarrow t \bar{b}$. It then sharply decreases to around 2.5$\fb$ for $m_{H^+}=$ 1 TeV. However, the fact that the decay $H \rightarrow W^{\pm}H^{\mp}$ has more phase space and the coupling 
$H H^{\pm} W^{\mp}$ is maximized when $\sin(\beta - \alpha)\approx 1$ leads to a significant branching ratio for the decay $H \rightarrow H^{\pm}W^{\mp}$. Additionally, one can see that the decay mode $H^{+} \rightarrow \tau^+ \nu$ can also dominate in the case of large $\tan\beta$.

Additionally, it can be noted from Fig.\ref{fig:fig6} and Fig.\ref{fig:fig10} that, in type X, cross section for $\mu^+ \mu^- \to  t \bar{b} H^-$ is the same as the one  $e^+ e^- \to  t \bar{b} H^-$ at $e^+e^-$ colliders for $\tan \beta \le $ 40. Beyond this point, with higher $\tan \beta$, an increase in the cross section is observed at $\mu^+ \mu^-$ collider compared to $e^+e^-$ collider. Because of the muon Yukawa coupling (see Eq.\ref{eq:trigonometricexp}), conversely, in Type II, the cross sections are identical at both colliders.
 
 For the 2HDM type II, since LHC data enforce $\tan\beta$ to be not too large, the results for $\mu^+ \mu^- \to t\bar{b} H^-$  are quite similar to the case of $e^+e^-$ with 3 TeV 
 and are not shown here. 

The dominance of one of these branching ratios means the suppression of the others including the fermionic decay of the charged Higgs. In 2HDM type X, one can suppress the channel $H^+\to t\bar{b}$ by taking $\tan\beta$ relatively large. 

\begin{figure}[!ht]
\centering
\includegraphics[width=0.244\textwidth]{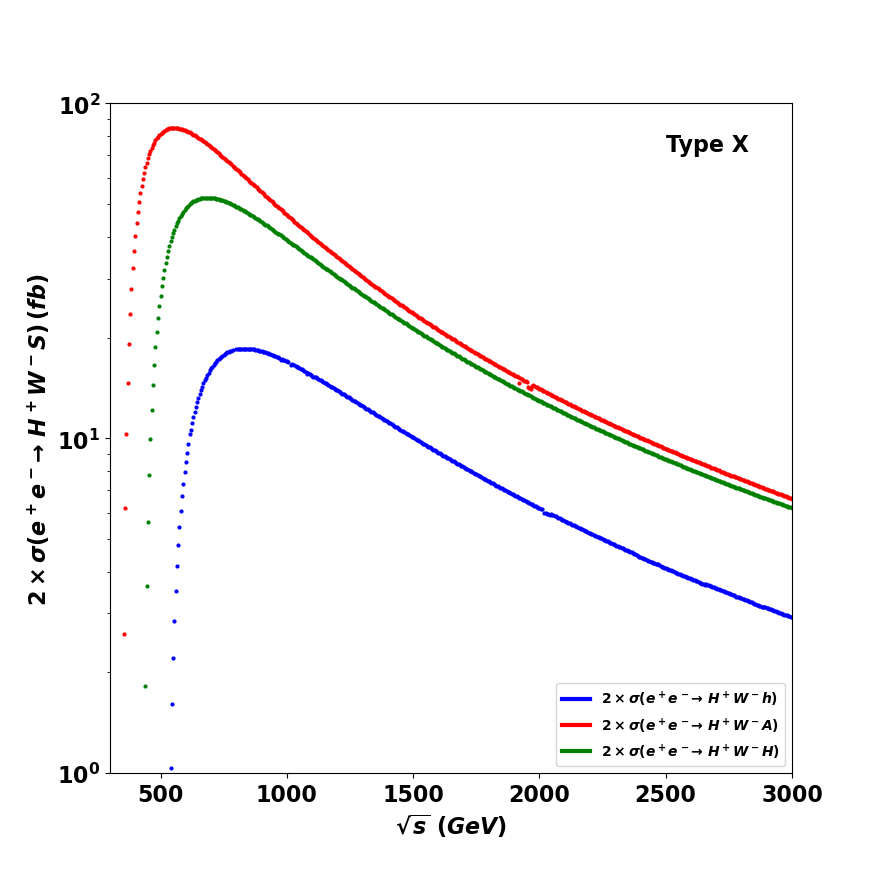}
\includegraphics[width=0.244\textwidth]{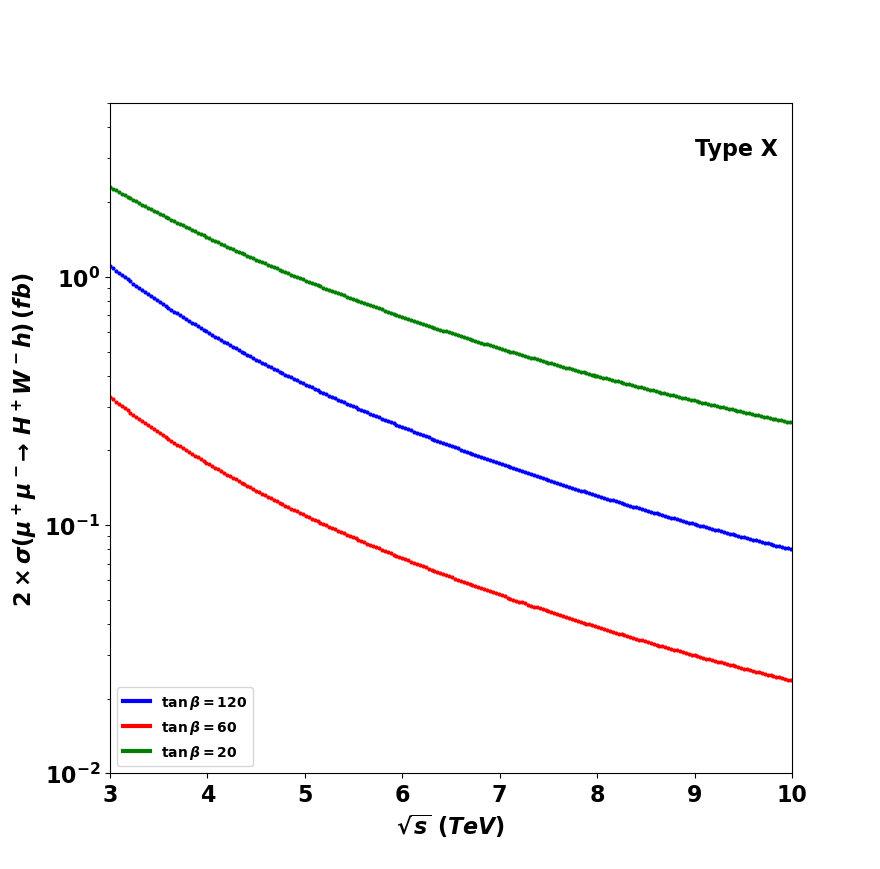}
\includegraphics[width=0.244\textwidth]{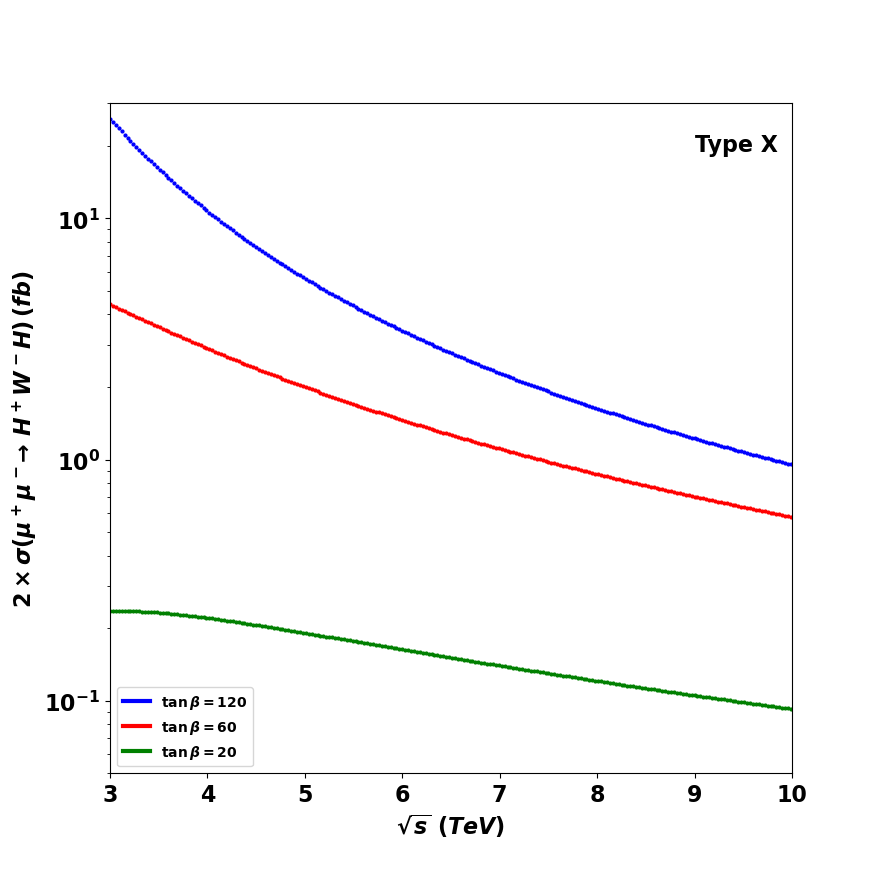}
\includegraphics[width=0.244\textwidth]{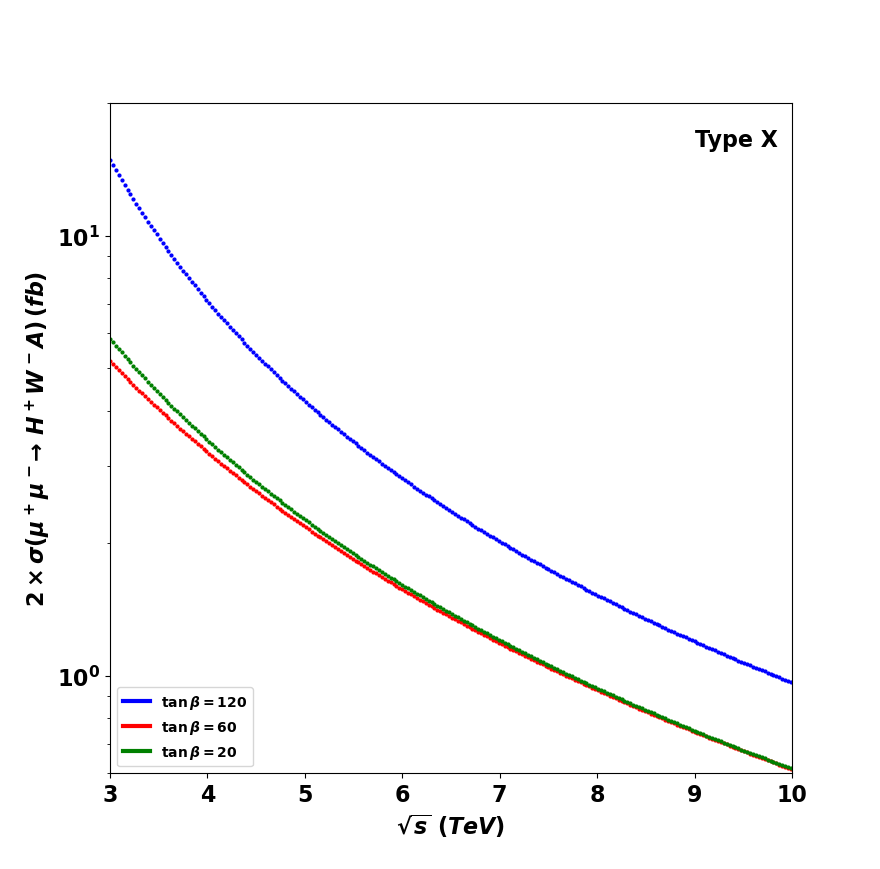}
\caption{The cross-section as a function of the center of mass energy $\sqrt{s}$ for $e^+e^- \to W^\pm H^\mp S_i$ $(i=h,\,H,\,A)$ within the Type-X 2HDM at the ILC (upper panel) and MuC (lower panel). At MuC, the cross section for each process is exhibited for three values of $\tan\beta$.}
\label{fig:fig11}
\end{figure}

The production cross sections in both ILC/CLIC and MuC for the centre of mass energies ranging respectively as $[0.3,\,3]$ (TeV) and $[3,\,10]$ (TeV) are given in Fig.\ref{fig:fig11}. Within $e^-e^+$ colliders, the cross sections for all processes show some short enhancements at low energies as stated in the upper panel, before decreasing with $\sqrt{s}$. However, it is interesting to see in lower panel that in muon collider, unlike ILC/CLIC, the cross sections $\sigma(e^+e^- \to W^\pm H^\mp S_i)$ scale as $s^{-1}$, and dependence of $\tan\beta$ expresses itself inevitably in such scaling.

\section{Signal-Background analysis for [$\tau \nu$][$\tau \nu$] AND [$Wb\bar{b}$][$Wb\bar{b}$]}
\label{section5}
\subsection{Collider Analysis}
\label{section3subsec4}

In muon collider, we undertake a comprehensive investigation of the singly charged Higgs boson phenomena through the application of both cut-based and multivariate analyses, focusing on the two final states [$\tau \nu$][$\tau \nu$] and [$Wb\bar{b}$][$Wb\bar{b}$]  arising from  $\mu^+ \mu^- \rightarrow \tau^+ \nu_\tau H^{-} $ and $\mu^+ \mu^- \rightarrow t\bar{b}  H^{-}$. To replicate the signal events and closely emulate real-world conditions, we employ simulation techniques. We initiate the generation of parton-level events using the software package \texttt{MadGraph5\_aMC\_v3.4.1} \cite{Alwall:2014hca}. These generated samples are then linked with \texttt{Pythia-8.20} \cite{Sjostrand:2007gs} to simulate fragmentation and showering effects. Subsequently, the events undergo simulation with the \texttt{Delphes-3.4.5} \cite{deFavereau:2013fsa} framework, which simulates the detector response. For this, we utilize the muon collider Detector TARGET model. To cluster jets, we apply the anti-kt algorithm \cite{Cacciari:2008gp} using \texttt{Delphes} with a jet radius parameter of R = 0.5. At the Delphes level, we impose initial conditions for b-jet candidates, requiring a minimum transverse momentum of p$_T >$ 20 GeV to satisfy acceptance and trigger criteria. Following this, we implement a b-tagging efficiency of approximately 70$\%$ and introduce mistag rates for charm or light quark jets being tagged as b-jets, taking into account variations in pseudorapidity and energy. For each channel, we adopt the benchmarks set in Table \ref{Bp1}. To assess observability, we evaluate the statistical significance (S) using the formula: 
\begin{eqnarray}
S=\sqrt{ \mathcal{L}} \frac{\sigma _s}{\sqrt{\sigma _s+\sigma _b}}
\end{eqnarray} 
where $\sigma _s$ and $\sigma_b$ are both the signal and background cross sections after all the cuts.

\begin{table}[h]
\setlength{\tabcolsep}{7pt}
\renewcommand{\arraystretch}{1.2}
\centering
\begin{tabular}{|c|c| c| c|c| c| c|c|c|}       
\hline \hline 
&signal& $m_h$  & $m_H$ & $m_A$ & $m_{H^{\pm}}$  &  $\tan \beta$  &  $\sin (\beta -\alpha)$&$m_{12}^2$  \\\hline\hline
BP1&$[\tau \nu][\tau \nu]$  & 125.09 & 138.03 & 279.92 & 101.36 & 29.07 & 0.997 &654.47\\ \hline
BP2& $[Wb\bar{b}][Wb\bar{b}]$ & 125.09 &  292.6 & 484.7  & 283.9  & 2.7    &0.997  &27225.16\\ 
\hline \hline
\end{tabular}
\caption{The description of our BPs.}\label{Bp1}
\end{table} 

\subsubsection{$\mu^+\mu^- \to \tau^+ \nu_{\tau} H^{-}$}
In this section, our attention will be directed towards producing charged Higgs bosons at a future muon collider. Specifically, we explore the $H^{\pm} \tau \nu$ final state, where the charged Higgs decays into $\tau \nu$, representing a process that aims to produce a pair of charged Higgs bosons
\begin{center}
$\mu^{+} \mu^{-} \rightarrow H^{-} \tau^+ \nu \rightarrow  \tau^- \nu \tau^+ \nu $
\end{center}
The signal and background parton-level events are required to satisfy the basic cuts $p_T^{j} > 25$ GeV and $|\eta_{j}| < 2.5$. We apply the criteria for charged lepton identification and typical photon isolation $I(P)= \frac{1}{p_{T}^{P}} \Sigma p_{T_{i}} < 0.01$. Additionally, we consider the $\tau$-tagging efficiency and mistagging rate presented by $\tau$ $P_{\tau \rightarrow \tau}$=0.85  and $P_{j \rightarrow \tau }$=0.02. Based on these signal characteristics, the main SM backgrounds include $ZZ$, $WW$, $t\bar{t}$, $Z/\gamma jj$, and $Wjj$.  To consider the background $Wjj$ more fully, we select hadronic-decaying tau $\tau_h$ comes from $W$ decay and the other $\tau_h$ from a jet misidentified as $\tau_h$. To enhance clarity, we provide a summary of the background processes and decay modes as shown below. In our calculations, we have included both the signal conjugate process $\mu^{+} \mu^{-} \rightarrow \tau^+ \nu H^{-}$ and the listed background conjugate processes. 

\begin{itemize}
\item $\mu^{+} \mu^{-} \rightarrow t\bar{t}$
\item $\mu^{+} \mu^{-} \rightarrow VV$  where $V=Z, \ W $
\item $\mu^{+} \mu^{-} \rightarrow Z/\gamma \ jj$ with $Z \rightarrow \tau \tau $ and $Z \rightarrow \nu \nu$
\item $\mu^{+} \mu^{-} \rightarrow Wjj$ where one $\tau_h$ comes from W decay and the other $\tau_h$ from a jet misidentified as $\tau_h$.
\end{itemize}

\begin{figure}[!ht]
\centering
\includegraphics[width=0.44\textwidth]{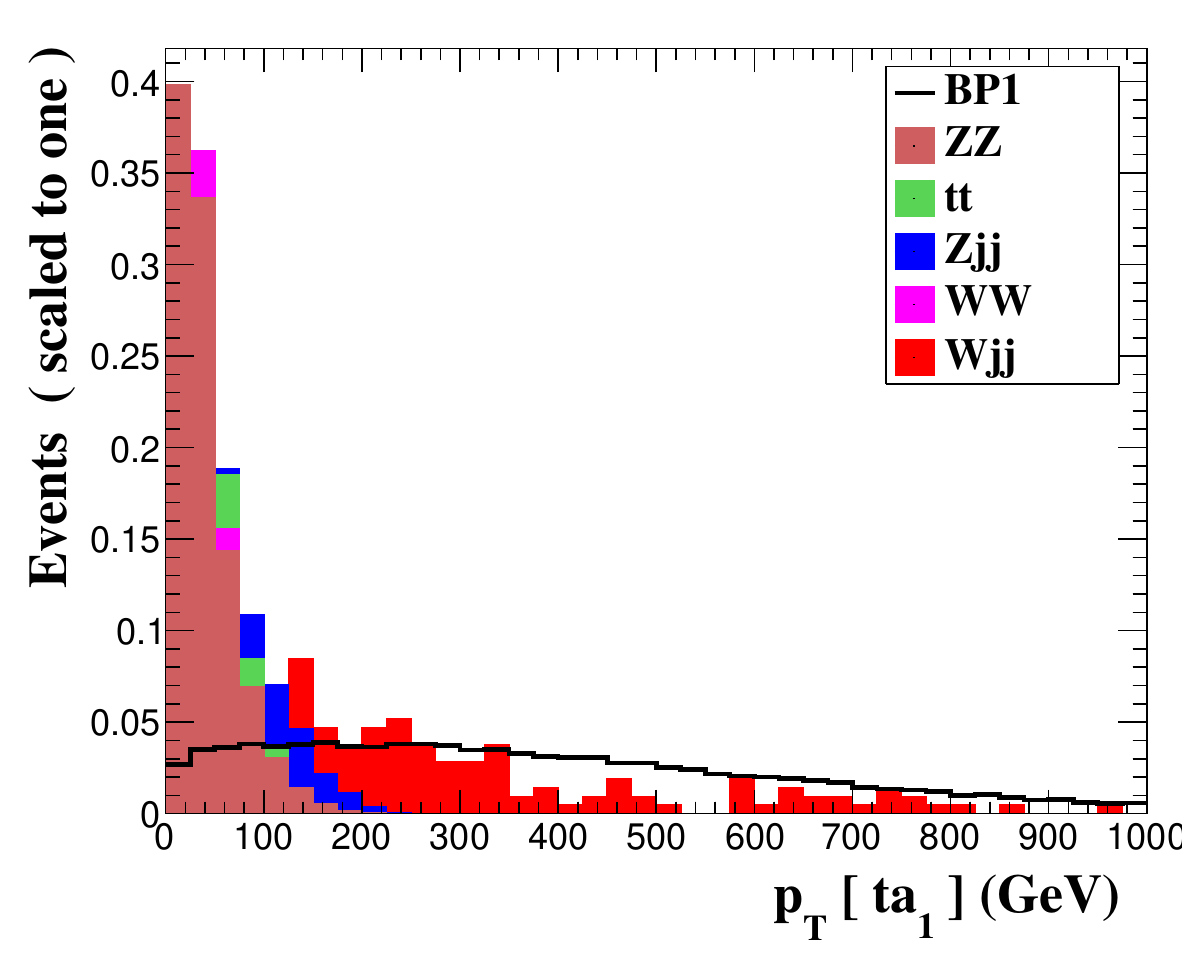}
\includegraphics[width=0.44\textwidth]{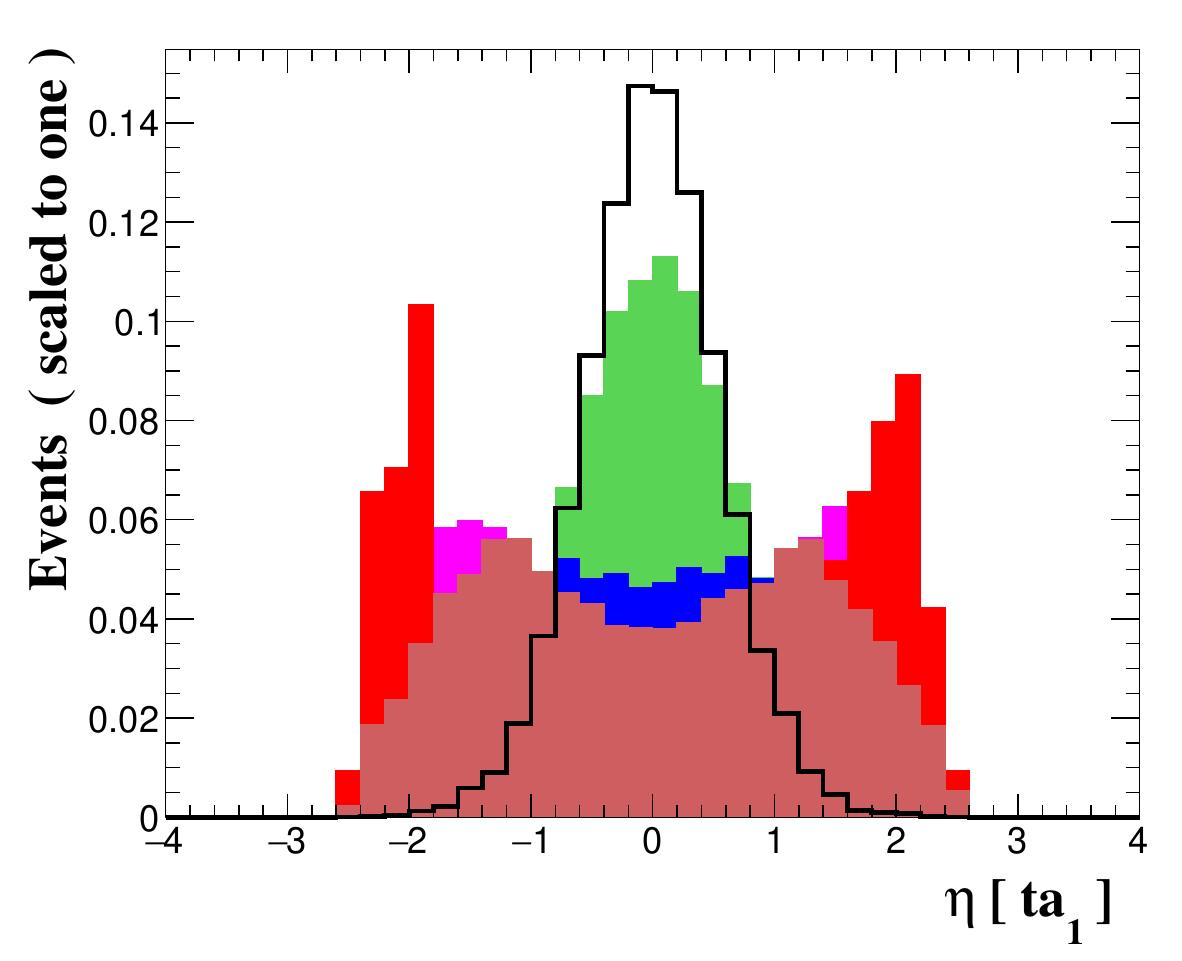}
\includegraphics[width=0.44\textwidth]{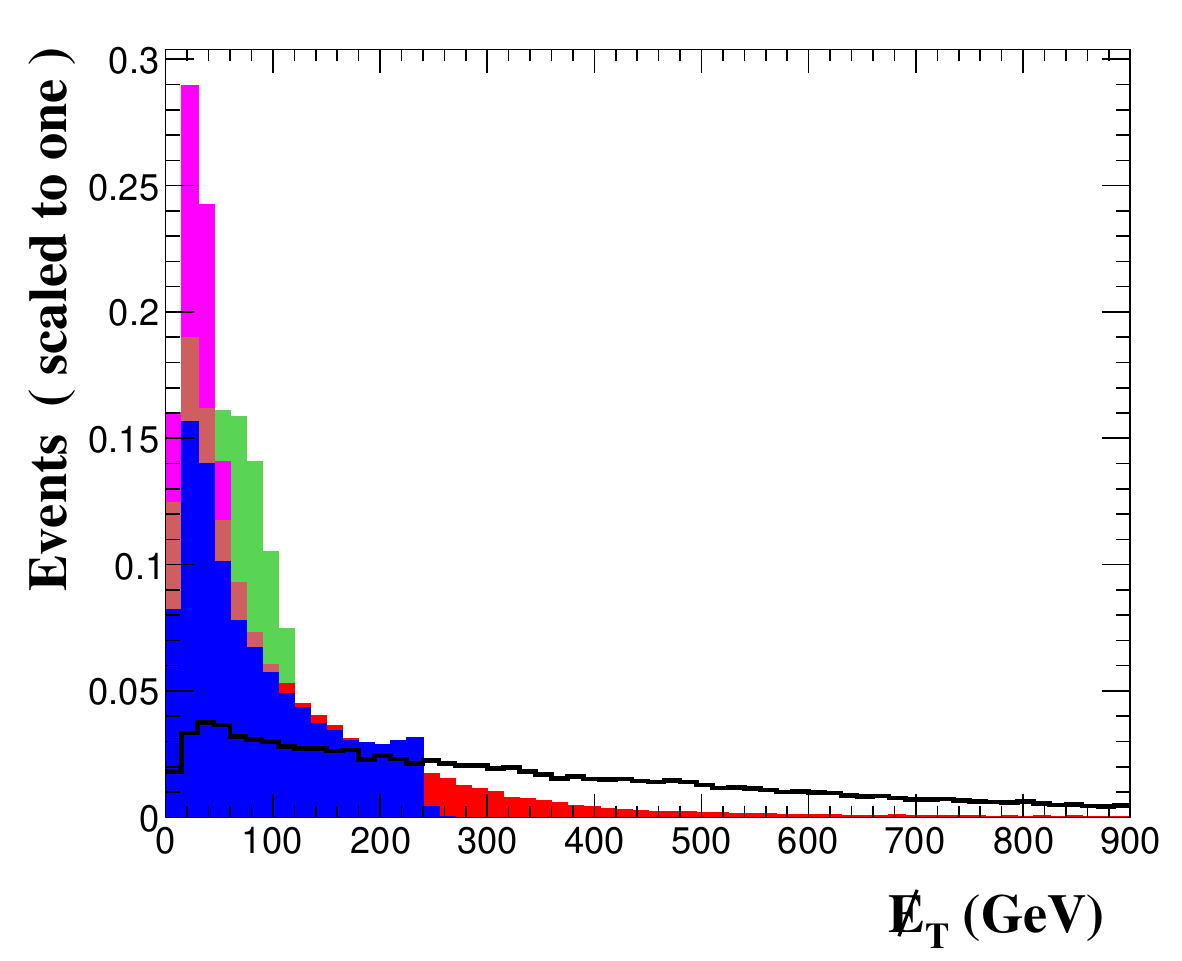}
\caption{Normalized kinematic distributions for the final state  $\tau^+ \nu \tau^-  \bar{\nu}$ about momentum transverse $p_T$[$\tau_1 $] ( left panel), the pseudorapidity of tau lepton $\eta$ [$\tau_1 $ ] (right panel) and the  missing transverse energy $\slashed{E}_T$ (lower panel) at $\sqrt{s}$=3 TeV  muon collider.} 
\label{fig:fig12}
\end{figure}

We apply restrictions on the counts of b-jets, with N(b)$\leq$ 1, a crucial selection to differentiate between the signal and backgrounds. This criterion alone leads to a reduction in the number of $t\bar{t}$ background by a factor of 4 without killing the signal. After applying the basic selection cuts to both signal and background events, we proceed to compute a range of kinematic distributions. We then present, in Fig.\ref{fig:fig12}, these distributions concerning momentum transverse $p_T$[$\tau_1 $] (top left panel),  the pseudorapidity of tau lepton $\eta$ [$\tau_1 $ ] (top right panel), and the last is the missing transverse energy $\slashed{E}_T$ (lower panel). The main backgrounds of $WW$, $Z/\gamma jj$ and $Wjj$ yield relatively are dominant. For this reason, we impose  $p_T$[$\tau_1 $] $>$ 150 GeV and  $-0.8< \eta [\tau_{1}]<0.8$. At the final stage of selection, we select the missing energy  $\slashed{E}_{T} > 260$  GeV, a total of 4890 signal events and 9 background events remain after the selection with the total integrated luminosity of $\mathcal{L}=3ab^{-1}$. Tab. \ref{TIII} shows cutflows on the cross sections (in$\fb$) for both the signal and the SM backgrounds at a center-of-mass energy of $\sqrt{s}=$3 TeV. With the number of signal and background events remaining, the significance appears very promising. Indeed, the muon collider Detector TARGET can explore the charged Higgs $H^{\pm}$ via the [$\tau \nu$][$\tau \nu$] final state. 

\begin{table}[]
\centering
\setlength{\tabcolsep}{5.pt}
\renewcommand{\arraystretch}{1}
\begin{tabular}{p{3cm}<{\centering}  p{1.5cm}<{\centering} p{1.4cm}<{\centering}p{1.4cm}<{\centering}  p{1.6cm}<{\centering} p{1.6cm}<{\centering} p{1.4cm}<{\centering} p{1.4cm}<{\centering} p{2cm}<{\centering} p{2cm}<{\centering} p{2cm}<{\centering}p{0cm}<{\centering}}
\hline\hline
\multirow{2}{*}{Cuts$\ \ \ \ \ \ \ \ \ \ \ \ \ \ \ $}& \multicolumn{1}{c}{Signal }& \multicolumn{3}{c}{~~Backgrounds}&  \\ \cline{2-2}  \cline{4-8}
&  $\text{BP}1$ &&$t\bar t $& $WW$  & $ZZ$ & $Wjj$&$Z/\gamma \ jj$\\
\hline\hline
Basic cut $\ \ \ \ \ \ \ \ $ &6.01  &&5.5 &61.67&1.55& 14&34.36\\
Tagger $\ \ \ \ \ \ \ \ \ \ \ $& 6.01 &&2.32 &61.17 &1.55 &13.89 & 34.3\\
Cut-1$\ \ \ \ \ \ \ \ \ \ \ \ \ \ $&2.53  && 0.0087 &0.17 &0.007&0.01 &0.35\\
Cut-2$\ \ \ \ \ \ \ \ \ \ \ \ \ \ $&1.63  && 0 &0 &0&0.003 &0\\
Total efficiencies& $27\%$ &&-&-& - &$0.02\%$&- \\
\hline\hline
\end{tabular}
\caption{The cut-flow chart of the cross section (in$\fb$) counts for both the signal and backgrounds in the $[\tau \nu ][\tau \nu] $ channel at the 3 TeV muon collider, with  our typical $\text{BP}1$ \label{TIII}.}
\end{table}

Summary of the cut schemes are offered as follow:
\begin{itemize}
\item			Trigger: $N(b)\leq 1$ 
\item			Cut-1:  $P_{T}[\tau_{1}] > 150 $ GeV   and    $-0.8< \eta [\tau_{1}]<0.8$ 
\item			Cut-2: $\slashed{E}_{T} > 260$  GeV    
\end{itemize}

\begin{figure}[h]
\centering
\includegraphics[width=0.5\textwidth]{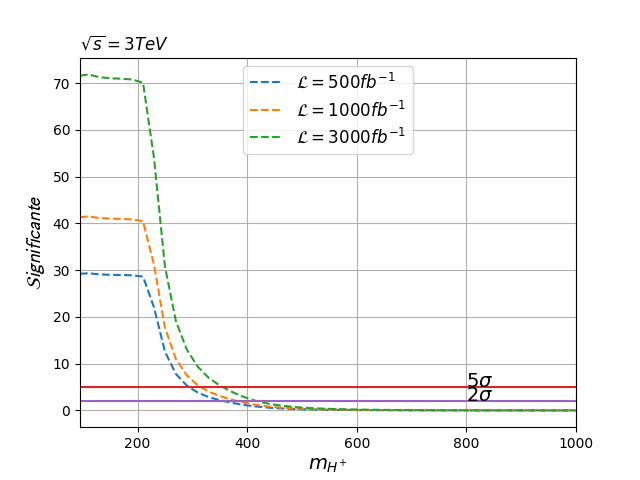}
\caption{The significance obtained for the process
$\mu^+ \mu^- \to \tau^+ \nu H^- \to \tau^+ \nu \tau^- \nu$ versus the
charged-Higgs mass at 3 TeV muon collider with
integrated luminosities of 500, 1000, and 3000 $\fb^{-1}$
with our benchmark point BP$_1$. The 2$\sigma$ and 5$\sigma$ significance levels are also indicated.}
\label{fig:fig13}
\end{figure}  

Fig.\ref{fig:fig13} illustrates the 5$\sigma$ and 2$\sigma$ limit capabilities in the significance-$m_{H^+}$ plane at $\sqrt{s}$ = 3 TeV for various integrated luminosities. The significance is shown as a function of $m_{H^+}$ for integrated luminosities of $\mathcal{L}$ = 500 $\fb^{-1}$, 1000 $\fb^{-1}$, and 3000 $\fb^{-1}$, where the red solid line denotes the 5$\sigma$ discovery and the purple solid line denotes the 2$\sigma$ exclusion. As observed, a significance of 5$\sigma$  can be achieved with a given integrated luminosity. The preferred region for discovering the charged Higgs boson at  $\sqrt{s}$ = 3 TeV is when $m_{H^+}$ ranges between approximately 280 GeV and 360 GeV. The prospects for significance appear highly promising. For $m_{H^+}$ values higher than 360 GeV, the significance is further reduced, and the 5$\sigma$ level cannot be reached. In conclusion, the provided integrated luminosities are crucial for achieving the 5$\sigma$ and 2$\sigma$ significance levels, which is essential for discovering the charged Higgs boson  $H^{\pm}$ through the process $\mu^+ \mu^- \to \tau^{+} \nu H^{-}$.

\subsubsection{$\mu^+\mu^- \to t \bar{b} H^{-}$}
The final state [$Wb\bar{b}$][$Wb\bar{b}$] is designed to target the production of a single of charged Higgs bosons, subsequently followed by $H^{+} \rightarrow t \bar{b}$

\begin{center}
$\mu^{+} \mu^{-} \rightarrow t \bar{b} H^{-}  \rightarrow  t \bar{b} \bar{t} b \rightarrow  W^+b\bar{b}W^-b\bar{b} \rightarrow 2l + 4b + \slashed{E}_T$
\end{center}
We take into consideration the benchmark point BP2 in Tab. \ref{Bp1}. We initiate our analysis by implementing acceptance cuts, which are applied to variables such as pseudorapidity ($\eta$), transverse momentum ($p_T$), and cone separation ($\Delta R$). These criteria aim at selecting the most pertinent events for subsequent analysis.

\begin{eqnarray}
p_{T}^{l} > 20\  GeV,\hspace{4mm} p_{T}^{b}> 25\ GeV , \hspace{4mm}|\eta| < 2.5 ,\hspace{4mm} \Delta R(l,l) > 0.4
\end{eqnarray}

The main SM backgrounds stem from top-pair production in association with a pair of b quarks denoted by $t\bar{t}b\bar{b}$ backgrounds. Additionally, top-pair production in association with the Standard model Higgs and with the Z boson $t\bar{t}V$, where $V=h,\ Z$, also contribute to the background; as well as production of Higgs boson and two Z bosons $ ZZh$, which are of minor consequence due to their smaller production cross sections. For clarity, we list these production processes and decay modes for both the signal and backgrounds as follows:
\begin{itemize}
\item $\mu^{+}\mu^{-} \rightarrow t\bar{t} b\bar{b} , (t \rightarrow  W^+ b, W^+  \rightarrow l^{+ } \nu_{l}),(\bar{t} \rightarrow  W^- \bar{b}, W^-  \rightarrow l^{-} \nu_{l})$
\item $\mu^{+}\mu^{-} \rightarrow t\bar{t} Z , (t \rightarrow  W^+ b, W^+  \rightarrow l^{+ } \nu_{l}),(\bar{t} \rightarrow  W^- \bar{b}, W^-  \rightarrow l^{-} \nu_{l}) , (Z \rightarrow b\bar{b})$
\item $\mu^{+}\mu^{-} \rightarrow t\bar{t} h , (t \rightarrow  W^+ b, W^+  \rightarrow l^{+ } \nu_{l}),(\bar{t} \rightarrow  W^- \bar{b}, W^-  \rightarrow l^{-} \nu_{l}) , (h \rightarrow b\bar{b})$
\item $\mu^{+}\mu^{-} \rightarrow ZZh , (Z  \rightarrow l^+ l^- ), (Z  \rightarrow b \bar{b}) ,(h  \rightarrow b \bar{b})$
\end{itemize}

\begin{figure*}[h]

\centering
\includegraphics[width=0.44\textwidth]{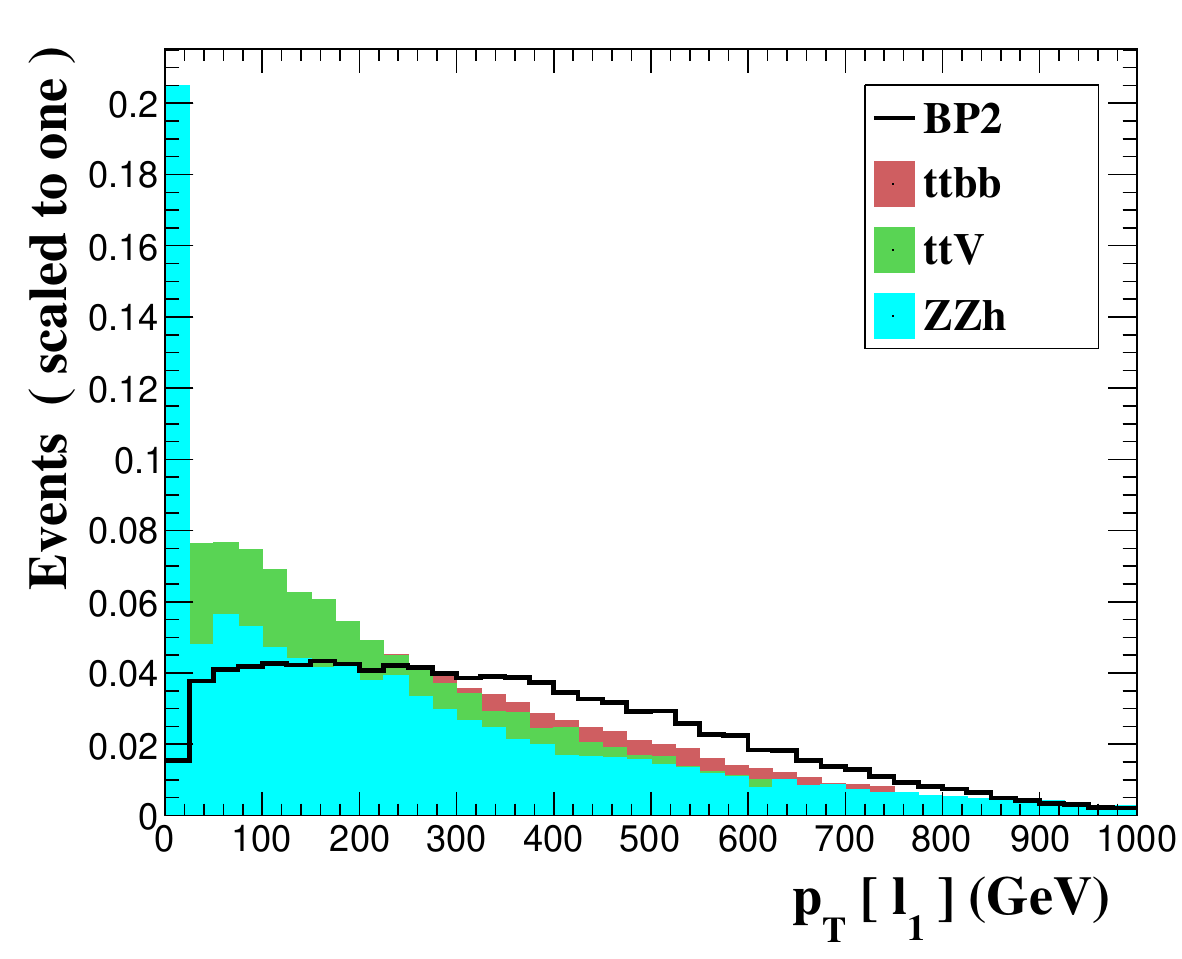}
\includegraphics[width=0.44\textwidth]{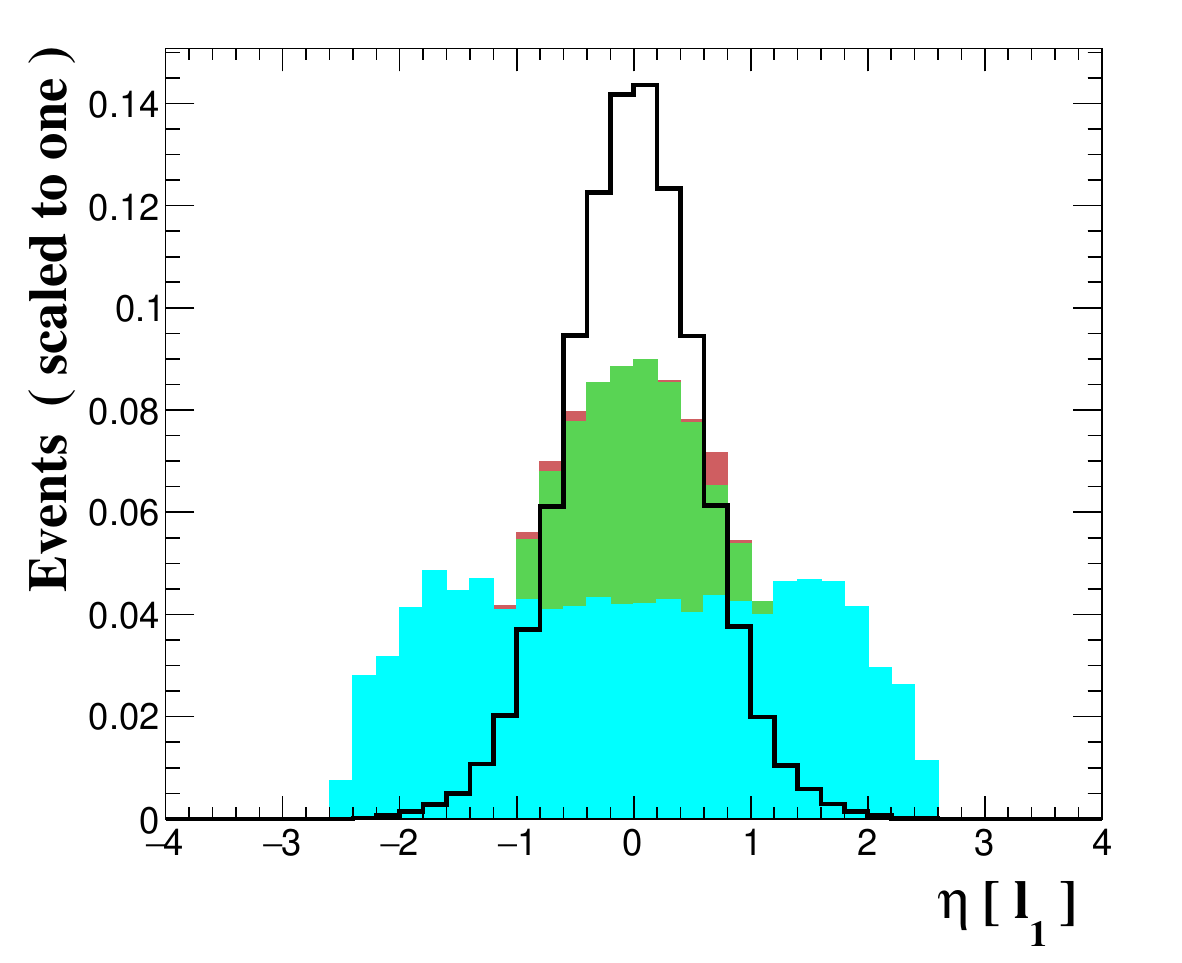}
\includegraphics[width=0.44\textwidth]{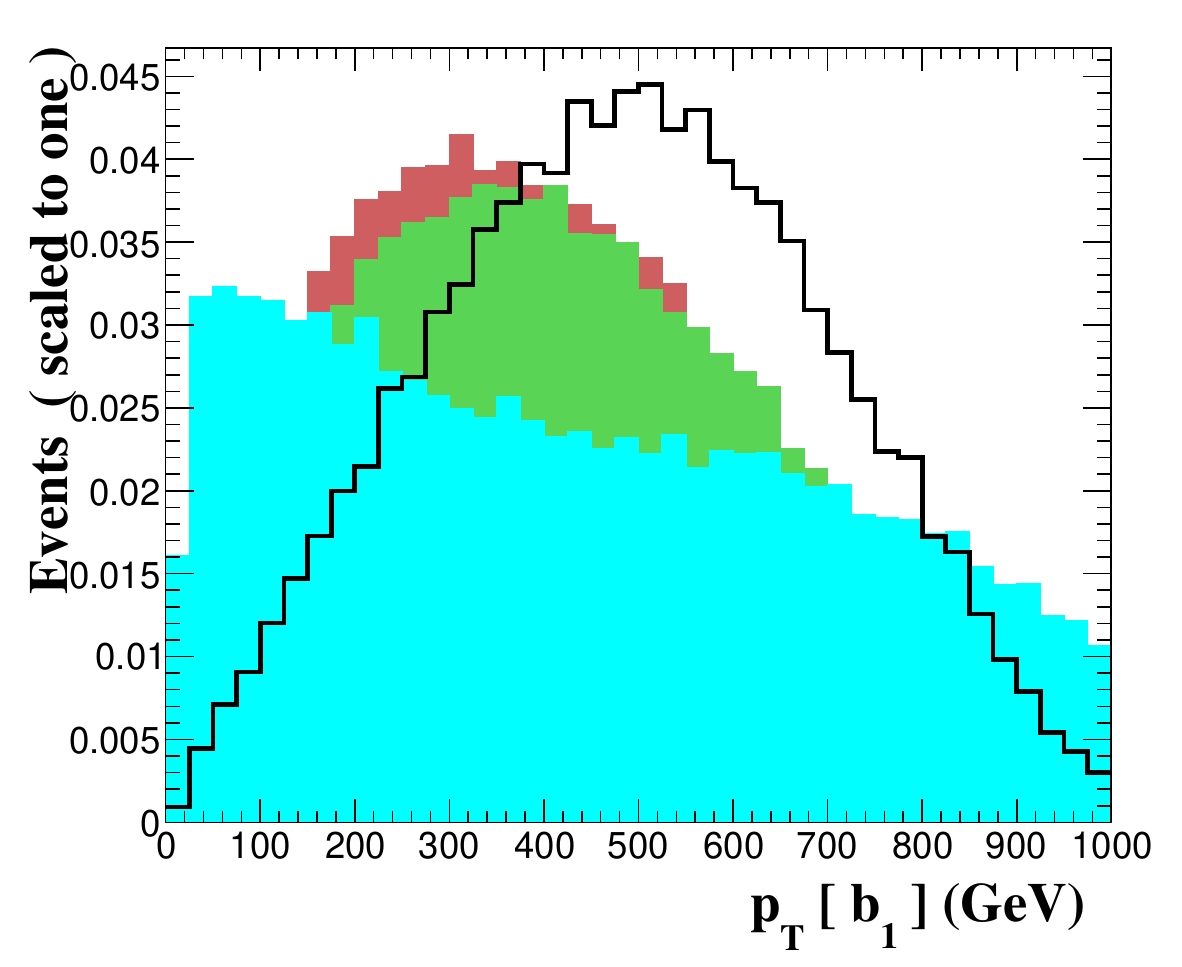}
\includegraphics[width=0.44\textwidth]{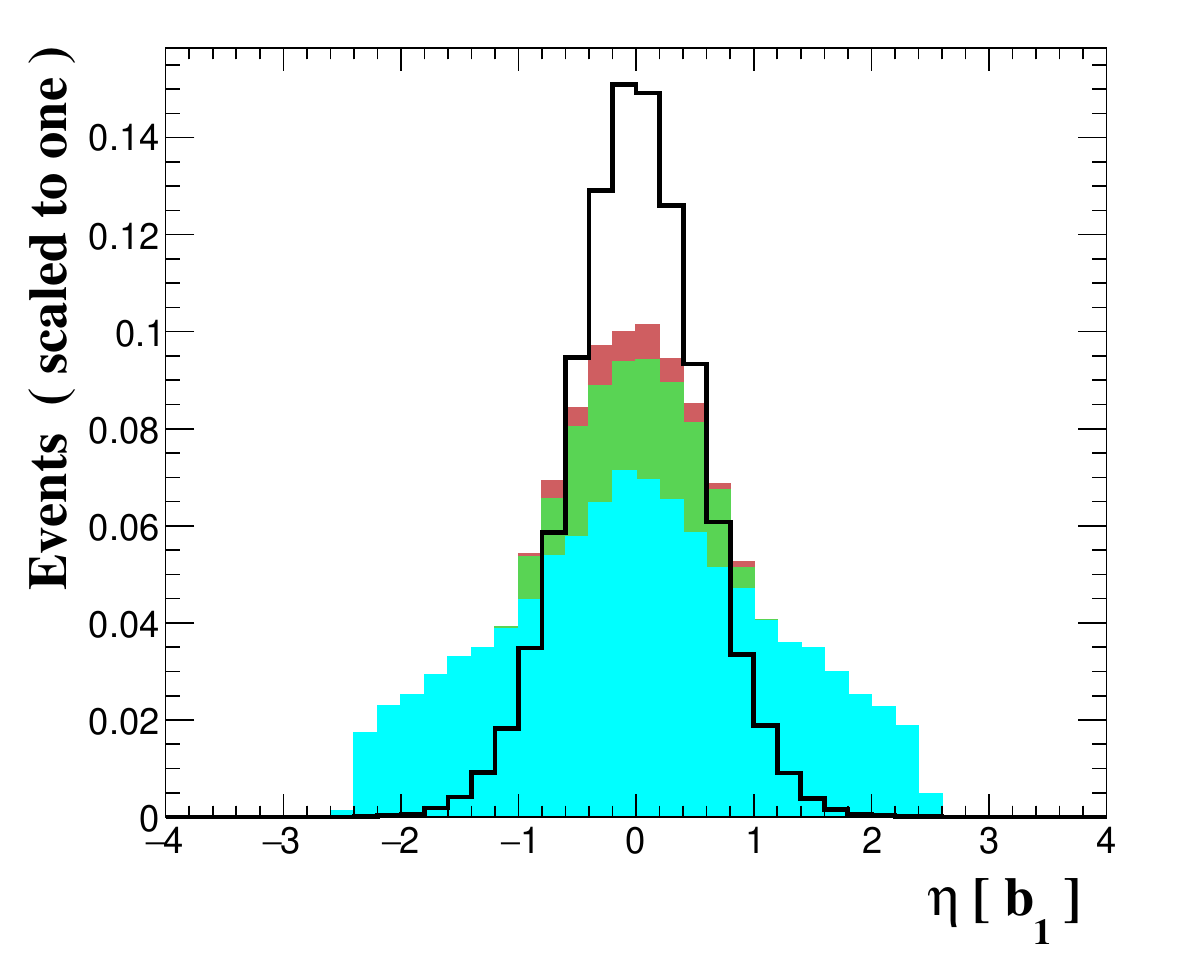}
\caption{Normalized kinematic distributions for the final state  $Wb\bar{b}Wb\bar{b}$ about transverse momentum of the lepton $p_T[l_1]$ (top left panel), the pseudorapidity of lepton $\eta$ [$l_1 $ ] (top right panel), transverse momentum of the b quark $p_T[b_1]$  (lower left panel), the last is the pseudorapidity of b quark $\eta$ [$b_1 $ ] (top right panel) (lower right panel) at $\sqrt{s}$=3 TeV muon collider.} 
\label{fig:fig14}
\end{figure*}

We show in Fig.\ref{fig:fig14} the kinematic distributions of the signal and backgrounds for the transverse momentum of the lepton $p_T[l_1]$ (top left panel), the pseudorapidity of lepton $\eta$ [$l_1 $ ] (top right panel), transverse momentum of the b quark $p_T[b_1]$  (lower left panel), and the pseudorapidity of b quark $\eta$ [$b_1$ ]  (lower right panel) at $\sqrt{s}$=3 TeV muon collider. To enhance the signal significance, we have implemented a set of selection cuts guided by the kinematic distributions.

We introduce a set of constraints aimed at refining the event pool, with particular emphasis (basic cut). An in-depth analysis of these distributions reveals that the main backgrounds of $t\bar{t}b\bar{b}$, $t\bar{t}h/Z$, and $ZZh$ hold substantial dominance. Consequently, we impose a stringent criterion the transverse momentum of the lepton and 
the pseudorapidity of  lepton $P_T [l_1] > 240 $GeV and $-0.8< \eta [l_{1}]<0.8\footnote{The leptons and b quarks are labeled by descending order of $p_T$, i.e, $p_T(\ell_1) > p_T(\ell_2)$.}$  going together for the detection of the
leptons, under this cut significantly reduces a substantial portion of the background.   Complementary to this, we impose "Cut-2" $P_T [b_1] > 400$ GeV and $-0.6< \eta [b_{1}]<0.6$. Applying this cut significantly reduces a substantial portion of the background as shown in Tab.\ref{CUTb}, making it easier to detect the b-quark.

\begin{table}[]
\centering
\setlength{\tabcolsep}{5.pt}
\renewcommand{\arraystretch}{1}
\begin{tabular}{p{3cm}<{\centering}  p{2.5cm}<{\centering} p{1.4cm}<{\centering}p{2.4cm}<{\centering}  p{1.6cm}<{\centering} p{2.6cm}<{\centering} p{1.4cm}<{\centering} p{2.4cm}<{\centering} p{3cm}<{\centering} p{3cm}<{\centering} }
\hline\hline
\multirow{2}{*}{Cuts$\ \ \ \ \ \ \ \ \ \ \ \ \ \ \ $}& \multicolumn{1}{c}{Signal }& \multicolumn{3}{c}{~~Backgrounds}&  \\ \cline{2-2}  \cline{4-6}
&  $\text{BP}2$ &&$t\bar t bb $& $t\bar t V$  & $ZZh$\\
\hline\hline
Basic cut $\ \ \ \ \ \ \ \ $ &0.185  &&0.007 &0.016&0.0012 \\
Cut-1$\ \ \ \ \ \ \ \ \ \ \ \ \ \ $&0.09 && 0.002 &0.004 &0.0002\\
Cut-2$\ \ \ \ \ \ \ \ \ \ \ \ \ \ $&0.05 && 0.0009&0.002 &0.0001 \\
Total efficiencies& $27\%$ &&12$\%$&12$\%$& 8$\%$  \\
\hline\hline
\end{tabular}
\caption{The cut-flow chart of the cross section (in$\fb$) counts for both the signal and backgrounds in the $[\tau \nu ][\tau \nu] bb$ channel at the 500GeV muon collider, with  our typical $\text{BP}2$ \label{CUTb}.}
\end{table}

Summary of the cut schemes are offered as follow:
\begin{itemize}
\item	 Cut-1:  $P_T [l_1] > 240$ GeV and $-0.8< \eta [l_{1}]<0.8$    
\item	 Cut-2:  $P_T [b_1] > 400$ GeV and $-0.6< \eta [b_{1}]<0.6$  
\end{itemize}

\begin{figure}[h]
\centering
\includegraphics[width=0.5\textwidth]{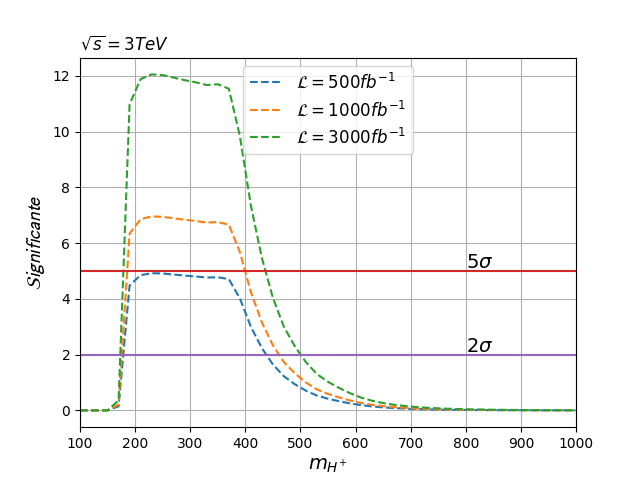}
\caption{
The significance obtained for the process
$\mu^+ \mu^- \to t \bar{b} H^-  \to 2l + 4b + \slashed{E}_T$ versus the
charged-Higgs mass at 3 TeV muon collider with integrated luminosities of 500, 1000, and 3000 $\fb^{-1}$ with our benchmark point BP$_1$. The 2$\sigma$ and 5$\sigma$ significance levels are also indicated.}
\label{fig:fig15}
\end{figure}  

Fig.\ref{fig:fig15} illustrates the 5$\sigma$ and 2$\sigma$ limit capabilities in the significance-$m_{H^+}$ plane at $\sqrt{s}$ = 3 TeV for various integrated luminosities. The significance is presented as a function of $m_{H^+}$ for integrated luminosities of $\mathcal{L}=$ 500 $\fb^{-1}$, 1000 $\fb^{-1}$, and 3000 $\fb^{-1}$. A 5$\sigma$ significance can be achieved with an integrated luminosity of  1000 $\fb^{-1}$ and 3000 $\fb^{-1}$. The most favorable range for discovering the charged Higgs boson at a $\sqrt{s}$ = 3 TeV muon collider is between approximately 100 GeV and 250 GeV. However, a 2$\sigma$  significance is achievable with integrated luminosities of 500 $\fb^{-1}$. This lower integrated luminosity is less promising, as the significance becomes too small to achieve the 5$\sigma$ level for the available charged Higgs mass. In conclusion, integrated luminosities higher than 500 $fb^{-1}$ are crucial for achieving both 5$\sigma$  and 2$\sigma$  significance levels, particularly for the discovery of the charged Higgs boson $H^{\pm}$ through the process $\mu^+ \mu^- \to t \bar{b} H^-$. 

\section{Conclusions}
\label{sec:conclusion}
Within the context of 2HDM, we have studied the singly charged Higgs production at the next generation of lepton colliders. We have examined the associated production of a singly charged Higgs with a gauge boson and with an additional neutral Higgs: $\ell^+ \ell^- \to W^\mp H^\pm S$, $S=h,\,H,\,A$ and $\ell^+ \ell^- \to W^\pm H^\mp  Z$., as well as with fermions: $\ell^+ \ell^- \to \tau^+ \nu H^-$ and $\ell^+ \ell^- \to t \bar{b} H^-$. We report results for all processes at the $e^+e^-$ collider with a center of mass energy of 500 GeV, 1 TeV, and 3 TeV, and at the muon collider with a center of mass energy of 3 TeV.\\
The analysis was conducted considering both experimental and theoretical restrictions, including multiple B physics measurements and LHC Higgs searches. For $\tan\beta< 40$,  we have found that in the 2HDM type X  the cross sections for $\ell^+ \ell^- \to \tau^+ \nu H^-$ and $\ell^+ \ell^- \to t \bar{b} H^-$ are rather small of the order a few$\fb$ both for $e^+e^-$ and muon colliders. While for large $\tan\beta>40$, in the case of  $\mu^+ \mu^- \to \tau^+ \nu H^-$ both charged Higgs and neutral Higgs couplings to muon could give a $\tan^6\beta$ factor to the amplitude square which could enhance the cross section with more than one order of magnitude compared to $e^+e^-$ results.\\
We have also shown that the single production of $H^\pm$ with a $W$ gauge boson and a neutral Higgs ($h,\,H$ or $A$) at $e^+e^-$ collider could be substantial in some cases. We explain that the cross section for $e^+ e^- \to W^\pm H^\mp h $ is rather small compared to the other ones: $e^+ e^- \to W^\pm H^\mp  H$ and  $e^+ e^- \to W^\pm H^\mp  A$ which could reach 10$\fb$ at $\sqrt{s}=$ 500 GeV and 1 TeV.  However, at very high energy with $\sqrt{s}=$ 3 TeV, we have shown that the cross sections for $e^+ e^- \to W^\pm H^\mp  H$ and  $e^+ e^- \to W^\pm H^\mp  A$  are greater than 0.1$\fb$ in quite a wide range for charged Higgs mass which is due to the presence of t-channel contribution. Similar to the associated production of $H^\pm$ with fermions, the associated production with W and neutral Higgs also benefits from a large $\tan\beta$ effect at the muon collider which is due to the contribution of neutral and charged Higgs. 

If the charged Higgs is too heavy to be produced in pairs in future lepton colliders, the single production in association with either fermions or bosons would be the only means to look for direct evidence for it. The smallness of the cross section would require, however, a very high luminosity option.

\subsection*{Acknowledgments}
This work is supported by the Moroccan Ministry of Higher Education and Scientific Research MESRSFC and CNRST: Projet PPR/2015/6. K.C. is supported in part by the National Science and Technology Council of Taiwan under the grant numbers MoST 110-2112-M-007-017-MY3 and 113-2112-M-007-041-MY3. AA would like to thank the Department of Physics and CTC, National Tsing Hua University, Taiwan for their hospitality during the course of this work. The authors would like to thank M. Krab for useful discussions.

\bibliographystyle{JHEP}
\bibliography{bibliography}
\end{document}